\documentclass[a4paper,11pt]{article}
\pdfoutput=1 

\usepackage{jcappub} 

\usepackage[T1]{fontenc} 
\usepackage{graphicx}
\graphicspath{{figures/}}
\usepackage[utf8]{inputenc}
\usepackage{orcidlink}
\usepackage{amsmath}
\usepackage{amsfonts}
\usepackage{amssymb}
\usepackage{enumitem}
\usepackage{multirow}
\usepackage{CJKutf8}
\usepackage{color}
\usepackage[capitalise]{cleveref}
\usepackage{acro}
\usepackage{adjustbox}
\usepackage{array}
\usepackage{natbib}
\usepackage{dblfnote}
\DFNalwaysdouble
\usepackage{slashed}
\usepackage{dcolumn}
\usepackage{hyperref}
\usepackage{geometry}
\usepackage{subfig} 
\usepackage{overpic}
\usepackage[most]{tcolorbox} \usepackage{lipsum}
\usepackage{inputenc}
\usepackage{caption}
\def\({\left(}
\def\){\right)}
\def\[{\left[}
\def\]{\right]}
\def\be{\begin{eqnarray}}
\def\ee{\end{eqnarray}}

\DeclareAcronym{GW}{
  short = GW ,
  long = gravitational wave ,
  short-plural = s 
}
\DeclareAcronym{LIGO}{
  short = LIGO ,
  long = Laser Interferometer Gravitational-wave Observatory ,
  short-plural = 
}

\DeclareAcronym{BBN}{
  short = BBN ,
  long = Big Bang Nucleosynthesis  ,
  short-plural = 
}

\DeclareAcronym{LISA}{
  short = LISA ,
  long = Laser Interferometer Space Antenna ,
  short-plural =  
}
\DeclareAcronym{SKA}{
  short = SKA ,
  long = Square Kilometre Array ,
  short-plural =  
}  
  
\DeclareAcronym{SNR}{
	short = SNR ,
	long = signal-to-noise ratio ,
	short-plural = 
}

\DeclareAcronym{PTA}{
	short = PTA ,
	long = pulsar timing array ,
	short-plural = 
}

\DeclareAcronym{FLRW}{
  short = FLRW ,
  long = Friedmann-Lemaitre-Robertson-Walker ,
  short-plural =  
}

\DeclareAcronym{SIGW}{
	short = SIGW ,
	long = scalar induced gravitational wave ,
	short-plural =  s
}

\DeclareAcronym{PBH}{
	short = PBH ,
	long = primordial black hole ,
	short-plural =  s
}

\DeclareAcronym{TSIGW}{
	short = TSIGW ,
	long = tensor-scalar induced gravitational wave ,
	short-plural =  s
}

\DeclareAcronym{PGW}{
	short = PGW ,
	long = primordial gravitational wave ,
	short-plural =  s
}

\DeclareAcronym{CMB}{
	short = CMB ,
	long = cosmic microwave background ,
	short-plural =  
}
\DeclareAcronym{DM}{
	short = DM ,
	long = dark matter ,
	short-plural =  
}

\DeclareAcronym{SGWB}{
	short = SGWB ,
	long = stochastic gravitational	wave background ,
	short-plural =  s
}

\DeclareAcronym{LSS}{
	short = LSS ,
	long = large-scale structure ,
	short-plural =  
}

\DeclareAcronym{BAO}{
	short = BAO ,
	long = baryon acoustic oscillations ,
	short-plural = 
}

\DeclareAcronym{RD}{
	short = RD ,
	long = radiation-dominated ,
	short-plural =  
}

\DeclareAcronym{SMBHB}{
  short = SMBHB ,
  long = supermassive black hole binary ,
  short-plural = s
}

\DeclareAcronym{KDE}{
  short = KDE ,
  long = kernel density estimator ,
  short-plural = s
}

\title{\boldmath Can tensor-scalar induced GWs dominate PTA observations ? }

\author[a]{Di Wu\orcidlink{0000-0001-7309-574X},}
\author[b,1]{Jing-Zhi Zhou\note{Corresponding author.} \orcidlink{0000-0003-2792-3182},}
\author[c,d]{Yu-Ting Kuang\orcidlink{0000-0002-7431-4454},}
\author[b]{Zhi-Chao Li\orcidlink{0009-0005-7984-2626},}
\author[c,d]{Zhe Chang\orcidlink{0000-0002-9720-803X},}
\author[a,d,e,1]{Qing-Guo Huang\orcidlink{0000-0003-1584-345X}}

\affiliation[a]{School of Fundamental Physics and Mathematical Sciences, Hangzhou Institute for Advanced Study, UCAS, Hangzhou 310024, China}
\affiliation[b]{Center for Joint Quantum Studies and Department of Physics,
School of Science, Tianjin University, Tianjin 300350, China}
\affiliation[c]{Institute of High Energy Physics, Chinese Academy of Sciences, Beijing 100049, China}
\affiliation[d]{University of Chinese Academy of Sciences, Beijing 100049, China}
\affiliation[e]{CAS Key Laboratory of Theoretical Physics, Institute of Theoretical Physics, Chinese Academy of Sciences}

\emailAdd{wudi@ucas.ac.cn}
\emailAdd{zhoujingzhi@tju.edu.cn}
\emailAdd{kuangyt@ihep.ac.cn}
\emailAdd{lizc@tju.edu.cn}
\emailAdd{changz@ihep.ac.cn}
\emailAdd{huangqg@itp.ac.cn}

\abstract{Observational constraints on small-scale primordial gravitational waves are considerably weaker than those on large scales. We focus on scenarios with significant primordial gravitational waves and curvature perturbations on small scales, studying the energy density spectrum of the second-order \acp{TSIGW}. By leveraging current data from  \ac{CMB}, \ac{BAO}, and \ac{PTA}, combined with the \ac{SNR} analysis of \ac{LISA}, we can investigate how tensor-scalar induced gravitational waves affect observations on various scales, thus constraining the parameter space for primordial gravitational waves and curvature perturbations. 
The Bayes factor analysis suggests that \acp{TSIGW}$+$\acp{PGW} might be more likely to dominate current \ac{PTA} observations compared to \ac{SMBHB}.
}

\begin{document}
\maketitle
\flushbottom

\section{Introduction}\label{sec:1.0}
Current cosmological observations indicate our universe is homogeneous and isotropic on large scales, as described by the \ac{FLRW} metric \cite{Baumann:2022mni}. Deviations from isotropy and homogeneity are described by cosmological perturbations in the \ac{FLRW} spacetime, and according to its symmetries, these perturbations are decomposed into scalar, vector, and tensor perturbations \cite{Malik:2008im,Wang:2013zva}. These types of cosmological perturbations play essential roles in cosmic evolution. Perturbations generated during inflation are known as primordial perturbations \cite{Baumann:2009ds}. As the initial conditions for the subsequent evolution of the universe, primordial perturbations influence all later cosmological processes.  

The cosmological observations allow us to constrain these primordial perturbations on different scales. As vector perturbations decay with $1/a^2$, we consider primarily primordial curvature (scalar) and primordial tensor perturbations (primordial gravitational waves) \cite{Lu:2008ju}. On large scales ($\gtrsim$1 Mpc), the amplitude of primordial curvature perturbations $A_{\zeta}$ is tightly constrained by \ac{CMB} and \ac{LSS} observations to around $A_{\zeta}\sim 2\times 10^{-9}$ \cite{Planck:2018vyg,Planck:2015fie,Planck:2015sxf}. The tensor-to-scalar ratio $r=A_{h}/A_{\zeta}$ is currently constrained below 0.06 \cite{Planck:2019nip, Abdalla:2022yfr,Perivolaropoulos:2021jda,Planck:2018jri}. However, on smaller scales ($\lesssim$1 Mpc), constraints on primordial curvature and tensor perturbations are much weaker \cite{Bringmann:2011ut,Carr:2020gox}. Current cosmological observations allow for large primordial curvature perturbations and \acp{PGW} on small scales. Large-amplitude primordial perturbations on small scales can generate significant observational effects after re-entering the horizon following inflation, leading to higher-order induced \acp{GW}. If we neglect \acp{PGW} on small scales and only consider large-amplitude primordial curvature perturbations, the resulting higher-order \acp{GW} are known as \acp{SIGW} \cite{Domenech:2021ztg,Matarrese:1997ay,Mollerach:2003nq,Ananda:2006af,Baumann:2007zm,Zhang:2022dgx,Mangilli:2008bw,Saga:2014jca,Papanikolaou:2021uhe,Li:2023qua,Chen:2022qec,Zhou:2021vcw,Chang:2022nzu,Chang:2022dhh,Chang:2023vjk,Chang:2023aba,Iovino:2024sgs,Lu:2024dzj}. When large primordial curvature perturbations and \acp{PGW} are both considered, the higher-order \acp{GW} induced by both are referred to as \acp{TSIGW} \cite{Chang:2022vlv,Yu:2023lmo,Bari:2023rcw,Gong:2019mui,Picard:2023sbz}.  Since the \ac{TSIGW} is sourced by primordial  perturbations with large amplitude on small scales ($\lesssim$1 Mpc),  it can be used as a probe to detect small-scale primordial power spectra \cite{Inomata:2018epa}.  

In June 2023, the \ac{PTA} collaborations NANOGrav \cite{NANOGrav:2023gor,NANOGrav:2023hvm}, EPTA \cite{EPTA:2023fyk}, Parkes PTA \cite{Reardon:2023gzh}, and China PTA \cite{Xu:2023wog} reported evidence for an isotropic, stochastic background
of \acp{GW} within the nHz frequency range. There are many potential sources for the \acp{SGWB}. For standard astrophysical sources, the PTA signal is predominantly attributed to \acp{SMBHB} \cite{Middleton:2020asl,NANOGrav:2020spf}. In addition, the \ac{PTA} observations might also have a cosmological origin, such as first-order phase transitions \cite{Fujikura:2023lkn,Addazi:2023jvg,Jiang:2023qbm,Xiao:2023dbb,Wu:2023hsa,He:2023ado}, cosmic strings \cite{Ellis:2020ena,Ellis:2023tsl,Lazarides:2023ksx,Yamada:2023thl,Qiu:2023wbs}, and \acp{SIGW} \cite{Vaskonen:2020lbd,DeLuca:2020agl,Balaji:2023ehk,Franciolini:2023pbf,You:2023rmn,Zhao:2023joc,Wang:2023ost,Zhu:2023faa,Di:2017ndc,Iacconi:2024hmg,Tzerefos:2024rgb}. In this paper, we consider the scenario where second-order \acp{TSIGW} dominate current \acp{SGWB} observations. It requires the large primordial tensor perturbations on small scales ($\lesssim$1 Mpc) which can be realized by various early-universe models, such as  $G^2$-inflation \cite{Kobayashi:2012wm}, sound speed resonance \cite{Cai:2020ovp}, and spectator fields \cite{Gorji:2023ziy}. In this case, the second-order \acp{GW} are induced by scalar-scalar, scalar-tensor, and tensor-tensor coupling source terms.  The current  \ac{PTA} observations can be used as a probe to constrict the parameter space of small-scale primordial curvature and tensor power spectra. Furthermore,  \acp{TSIGW} and \acp{PGW} can contribute as an extra radiation component, impacting the large-scale cosmological observations \cite{Clarke:2020bil,Ben-Dayan:2019gll,Cang:2022jyc}. By combining current \ac{PTA} observations on small scales with  \ac{CMB} and \ac{BAO} observational data, we can investigate second-order \acp{TSIGW} and constrain the parameter space of small-scale \acp{PGW} and curvature perturbations. In addition, since the influence of \acp{PGW} on the second-order \acp{TSIGW} is primarily concentrated in the high-frequency region, we can rigorously analyze the \ac{SNR} of \ac{LISA} based on  theoretical results of \acp{TSIGW}, thus quantitatively describing the impact of \acp{PGW} on the total energy density spectrum of \acp{GW} in the high-frequency region.

This paper is organized as follows. In Sec.~\ref{sec:2.0}, we review the basic calculations of \acp{TSIGW}. In Sec.~\ref{sec:3.0}, we study the current total energy density spectra of \acp{GW}. The explicit expressions of the power spectra of \acp{TSIGW} are given in this section. In Sec.~\ref{sec:4.0}, we investigate the constraints on the parameter space of \acp{TSIGW} and the primordial power spectrum through combined experimental observations from \ac{CMB}, \ac{BAO}, \ac{PTA}, and \ac{LISA}. Finally, we summarize our results and give some discussions in Sec.~\ref{sec:5.0}.

\section{Equations of motion of TSIGWs}\label{sec:2.0}
In this section, we briefly review the equations of motion of second-order  \acp{TSIGW} \cite{Chang:2022vlv,Yu:2023lmo,Bari:2023rcw,Gong:2019mui}. The line element of perturbed spacetime in Newtonian gauge is expressed as \cite{Malik:2008im}
\begin{eqnarray}\label{eq:dS}
	\mathrm{d} s^{2}=a^{2}\left[-\left(1+2 \phi^{(1)}\right) \mathrm{d} \eta^{2}+\left(\left(1-2 \psi^{(1)}\right) \delta_{i j} +h^{(1)}_{ij}+\frac{1}{2}h^{(2)}_{ij}\right)\mathrm{d} x^{i} \mathrm{d} x^{j}\right] \ ,
\end{eqnarray}
where $\phi^{(1)}$ and $\psi^{(1)}$ are first-order scalar perturbations, $h^{(n)}_{ij}$$\left( n=1,2 \right)$ are $n$-order tensor perturbations. In the study of second-order \acp{TSIGW}, we need to consider large-amplitude \acp{PGW} on small scales. Therefore, unlike the usual second-order \acp{SIGW}, we retain the first-order tensor perturbation $h^{(1)}_{ij}$ in the metric perturbation given in Eq.~(\ref{eq:dS}). By substituting the metric perturbation in Eq.~(\ref{eq:dS}) into the Einstein field equations, we can obtain the equations of motion for the second-order \acp{TSIGW} \cite{Chang:2022vlv}
\begin{equation}\label{eq:h}
	\begin{aligned}
		h_{ij}^{(2)''}(\eta,\mathbf{x})+2 \mathcal{H} h_{ij}^{(2)'}(\eta,\mathbf{x})-\Delta h_{ij}^{(2)}(\eta,\mathbf{x})=-4 \Lambda_{ij}^{lm}\left( \mathcal{S}^{(2)}_{lm,\phi\phi}+\mathcal{S}^{(2)}_{lm,\phi h}+\mathcal{S}^{(2)}_{lm,hh}\right)  \ ,
	\end{aligned}
\end{equation}
where  $\mathcal{H}=a'/a$ is the conformal Hubble parameter, and $\Lambda_{ij}^{lm}$ is the decomposed operator to extract the transverse and traceless terms \cite{Zhou:2021vcw}.  In this paper, we consider the second-order \acp{TSIGW} during the \ac{RD} era, where $\mathcal{H}=1/\eta$ and $w=c_s^2=1/3$. The three types of source terms in Eq.~(\ref{eq:h}) can be expressed as
	\begin{eqnarray}
	\mathcal{S}^{(2)}_{lm,\phi\phi}(\eta,\mathbf{x})&=&\partial_{l} \phi^{(1)} \partial_{m} \phi^{(1)} +4 \phi^{(1)} \partial_{l} \partial_{m} \phi^{(1)}-\frac{1}{ \mathcal{H}}\left(\partial_{l} \phi^{(1)'} \partial_{m} \phi^{(1)}+\partial_{l} \phi^{(1)}  \partial_{m} \phi^{(1)'}\right) \nonumber\\
	&&-\frac{1}{ \mathcal{H}^{2}} \partial_{l} \phi^{(1)'} \partial_{m}  \phi^{(1)'} \ ,
	\label{eq:1} \\
	\mathcal{S}^{(2)}_{lm,\phi h}(\eta,\mathbf{x})&=& 10\mathcal{H}h^{(1)}_{lm}\phi^{(1)'}+3h^{(1)}_{lm}\phi^{(1)''}-\frac{5}{3}h^{(1)}_{lm}\Delta\phi^{(1)}-2\partial_{b}h^{(1)}_{lm}\partial^{b}\phi^{(1)}-2\phi^{(1)}\Delta h^{(1)}_{lm} \ ,
	\label{eq:2} \\
	\mathcal{S}^{(2)}_{lm,hh}(\eta,\mathbf{x})&=&\frac{1}{2}\left( -h^{b,(1)'}_{l} h^{(1)'}_{mb} +\partial_c h^{(1)}_{mb}\partial^c h^{b,(1)}_{l} -h^{bc,(1)}\partial_c \partial_m h^{(1)}_{lb}  -\partial_b h^{(1)}_{mc}\partial^c h^{b,(1)}_{l}  \right. \nonumber\\
	&&\left. +\frac{1}{2} h^{bc,(1)}\partial_l \partial_m h^{(1)}_{bc}+ h^{bc,(1)}\partial_c \partial_b h^{(1)}_{lm}- h^{bc,(1)}\partial_c \partial_l h^{(1)}_{mb} \right) \ .
	\label{eq:3} 
\end{eqnarray}
As shown in Eq.~(\ref{eq:1})--Eq.~(\ref{eq:3}), the source term $S_{lm,\phi \phi}$ is identical to the source term of the second-order \acp{SIGW} \cite{Domenech:2021ztg,Kohri:2018awv}. The source term $S_{lm,\phi h}$ is composed of the product of the first-order scalar perturbation $\phi^{(1)}$ and the first-order tensor perturbation $h^{(1)}_{lm}$. The source terms $S_{lm,hh}$ are composed of the first-order tensor perturbation $h^{(1)}_{lm}$. In momentum space, the solutions of first-order scalar and tensor perturbations are 
\begin{equation}\label{eq:pri}
	\psi(\eta,\mathbf{k}) = \phi(\eta,\mathbf{k})=\frac{2}{3}\zeta_{\mathbf{k}} T_\phi(|\mathbf{k}| \eta) \ , \
	h^{\lambda,(1)}(\eta,\mathbf{k}) = \mathbf{h}^{\lambda}_{\mathbf{k}} T_{h}(|\mathbf{k}| \eta) \ ,
\end{equation}
where $\zeta_{\mathbf{k}}$ and $\mathbf{h}^{\lambda}_{\mathbf{k}}$ are the primordial curvature and tensor perturbations, respectively. During the \ac{RD} era, the transfer functions $ T_\phi(|\mathbf{k}| \eta)$ and $T_{h}(|\mathbf{k}| \eta)$ in Eq.~(\ref{eq:pri}) are given by \cite{Kohri:2018awv}
\begin{equation}\label{eq:T}
	T_{\phi}(x)=\frac{9}{x^{2}}\left(\frac{\sqrt{3}}{x} \sin \left(\frac{x}{\sqrt{3}}\right)-\cos \left(\frac{x}{\sqrt{3}}\right)\right) \ , \  T_{h}(x)=\frac{\sin x}{x} \ ,
\end{equation}
where we have set $x\equiv |\mathbf{k}|\eta$. By substituting the analytical expressions of first-order perturbations in Eq.~(\ref{eq:pri}) and Eq.~(\ref{eq:T}) into  Eq.~(\ref{eq:1})-- Eq.~(\ref{eq:3}), we can derive the specific expressions for the three types of source terms of second-order \acp{TSIGW}. Using the Green's function approach to solve Eq.~(\ref{eq:h}), we obtain the expression for second-order \acp{TSIGW} as
\begin{equation}\label{eq:h10}
	\begin{aligned}
		h^{\lambda,(2)}(\eta,\mathbf{k})=\sum^{7}_{i=1} h_i^{\lambda,(2)}(\eta,\mathbf{k}) \ ,
	\end{aligned}
\end{equation}
where $h^{\lambda,(2)}(\eta,\mathbf{k})=\varepsilon^{\lambda, ij}(\mathbf{k})h_{ij}^{(2)}(\eta,\mathbf{k})$ and  $S^{\lambda,(2)}(\eta,\mathbf{k})=-\varepsilon^{\lambda, lm}(\mathbf{k})S_{lm}^{(2)}(\eta,\mathbf{k})$. And the $\varepsilon_{i j}^{\lambda}(\mathbf{k})$ is polarization tensor. The explicit expressions of $h_i^{\lambda,(2)}(\eta,\mathbf{k})$ $(i=1\sim 7)$ in Eq.~(\ref{eq:h10}) are given by
\begin{eqnarray}
	h_1^{\lambda,(2)}(\eta,\mathbf{k})&=&-\frac{4}{9}\int\frac{\mathrm{d}^3p}{(2\pi)^{3/2}}\varepsilon^{\lambda,lm}\left(\mathbf{k}\right)p_lp_mI^{(2)}_{1}\left( u,v,x \right)\zeta_{\mathbf{k}-\mathbf{p}}\zeta_{\mathbf{p}} \ ,
	\label{eq:h1}\\
	h_2^{\lambda,(2)}(\eta,\mathbf{k})&=&-\frac{2}{3}\int\frac{\mathrm{d}^3p}{(2\pi)^{3/2}}\varepsilon^{\lambda,lm}\left(\mathbf{k}\right) \varepsilon^{\lambda_1}_{lm}\left(\mathbf{p}\right)k^2 I^{(2)}_{2}\left( u,v,x \right) \zeta_{\mathbf{k}-\mathbf{p}}\mathbf{h}^{\lambda_1}_{\mathbf{p}} \ ,
	\label{eq:h2}\\
	h_3^{\lambda,(2)}(\eta,\mathbf{k})&=&-\int\frac{\mathrm{d}^3p}{(2\pi)^{3/2}}\varepsilon^{\lambda,lm}\left(\mathbf{k}\right) \varepsilon^{\lambda_1,b}_{l}\left(\mathbf{k}-\mathbf{p}\right)\varepsilon^{\lambda_2}_{bm}\left(\mathbf{p}\right)k^2I^{(2)}_{3}\left(u,v,x\right)\mathbf{h}^{\lambda_1}_{\mathbf{k}-\mathbf{p}}\mathbf{h}^{\lambda_2}_{\mathbf{p}} \ ,
	\label{eq:h3}\\
	h_4^{\lambda,(2)}(\eta,\mathbf{k})&=&-\int\frac{\mathrm{d}^3p}{(2\pi)^{3/2}}\varepsilon^{\lambda,lm}\left(\mathbf{k}\right) \varepsilon^{\lambda_1,bc}\left( \mathbf{k}-\mathbf{p} \right)\left(  2\varepsilon^{\lambda_2}_{mb}\left( \mathbf{p} \right)p_cp_l\right) \nonumber\\
 & & \times ~~I^{(2)}_4\left(u,v,x\right)\mathbf{h}^{\lambda_1}_{\mathbf{k}-\mathbf{p}}\mathbf{h}^{\lambda_2}_{\mathbf{p}} \ ,
	\label{eq:h4}\\
	h_5^{\lambda,(2)}(\eta,\mathbf{k})&=&-\int\frac{\mathrm{d}^3p}{(2\pi)^{3/2}}\varepsilon^{\lambda,lm}\left(\mathbf{k}\right) \varepsilon^{\lambda_1}_{mc}\left(\mathbf{k}-\mathbf{p}\right)\varepsilon^{\lambda_2}_{lb}\left(\mathbf{p}\right)\left( k-p \right)^bp^c \nonumber\\
 & & \times ~~I^{(2)}_{5}\left(u,v,x\right)\mathbf{h}^{\lambda_1}_{\mathbf{k}-\mathbf{p}}\mathbf{h}^{\lambda_2}_{\mathbf{p}} \ ,
	\label{eq:h5}\\
	h_6^{\lambda,(2)}(\eta,\mathbf{k})&=&\int\frac{\mathrm{d}^3p}{(2\pi)^{3/2}}\varepsilon^{\lambda,lm}\left(\mathbf{k}\right) \varepsilon^{\lambda_1}_{bc}\left(\mathbf{k}-\mathbf{p}\right)\varepsilon^{\lambda_2}_{lm}\left(\mathbf{p}\right)p^bp^cI^{(2)}_6\left(u,v,x\right)\mathbf{h}^{\lambda_1}_{\mathbf{k}-\mathbf{p}}\mathbf{h}^{\lambda_2}_{\mathbf{p}} \ ,
	\label{eq:h6} \\
	h_7^{\lambda,(2)}(\eta,\mathbf{k})&=&\int\frac{\mathrm{d}^3p}{(2\pi)^{3/2}}\varepsilon^{\lambda,lm}\left(\mathbf{k}\right) \varepsilon^{\lambda_1,bc}\left(\mathbf{k}-\mathbf{p}\right)\varepsilon^{\lambda_2}_{bc}\left(\mathbf{p}\right)\frac{p_lp_m}{2} I^{(2)}_{7}\left(u,v,x\right) \mathbf{h}^{\lambda_1}_{\mathbf{k}-\mathbf{p}}\mathbf{h}^{\lambda_2}_{\mathbf{p}} \ .
	\label{eq:h7}
\end{eqnarray}
Here, based on the specific form of the source term, we decompose the contributions of the source term  $\mathcal{S}^{(2)}_{lm,hh}$ into five parts: $h^{\lambda,(2)}_3 \sim h^{\lambda,(2)}_7$. We defined $|\mathbf{k}-\mathbf{p}|=u|\mathbf{k}|$ and $|\mathbf{p}|=v|\mathbf{k}|$. The kernel functions $I^{(2)}_i\left( u,v,x \right)$ in Eq.~(\ref{eq:h1})--Eq.~(\ref{eq:h7}) are given by
\begin{equation}\label{eq:I}
	\begin{aligned}
		I^{(2)}_i\left( u,v,x \right)=\frac{4}{k^2} \int_{0}^{x} \mathrm{d} \bar{x} \left( \frac{\bar{x}}{x}\sin\left( x-\bar{x} \right) f_i\left( u,v,\bar{x} \right)  \right) \ , \ (i=1 \sim 7) \ ,
	\end{aligned}
\end{equation}
where
\begin{eqnarray}
	f_1^{(2)}\left( u,v,x \right)&=&x^2uv\frac{\mathrm{d}}{\mathrm{d}(ux)} T_{\phi}\left( ux \right)\frac{\mathrm{d}}{\mathrm{d}(vx)}T_{\phi}\left( vx \right)+2ux T_{\phi}\left(vx \right)\frac{\mathrm{d}}{\mathrm{d}(ux)}T_{\phi}\left( ux \right)\nonumber\\
 &+&3T_{\phi}\left(ux \right)T_{\phi}\left( vx \right) \, ,
	\label{eq:f1}\\
	f_2^{(2)}\left( u,v,x \right)&=&\frac{10u}{x}T_{h}\left( vx \right)\frac{\mathrm{d}}{\mathrm{d}(ux)} T_{\phi}\left( ux \right)+3u^2T_{h}\left( vx \right)\frac{\mathrm{d}^2}{\mathrm{d}(ux)^2}T_{\phi}\left(ux \right)+\frac{5}{3}u^2 T_{\phi}\left( ux \right)T_{h}\left( vx \right) \nonumber\\
	&+&\left(1-v^2-u^2\right)T_{\phi}\left( ux \right)T_{h}\left( vx \right)+2v^2T_{\phi}\left( ux \right)T_{h}\left( vx \right) \, ,
	\label{eq:f2}\\
	f_{3}^{(2)}\left(u,v,x\right)&=&-\frac{1-u^2-v^2}{4}T_{h}\left( ux \right)T_{h}\left( vx \right)-\frac{uv}{2}\frac{\mathrm{d}}{\mathrm{d}(ux)}T_{h}\left( ux \right)\frac{\mathrm{d}}{\mathrm{d}(vx)}T_{h}\left( vx \right) \, ,
	\label{eq:f3} \\
	f_i^{(2)}\left( u,v,x \right)&=&\frac{1}{2}T_{h}\left( ux \right)T_{h}\left( vx \right) \, , \  (i=4,5,6,7)  
	\label{eq:f4} \, .
\end{eqnarray}
The analytic expressions of the second-order kernel functions $I^{(2)}_i$$\left(i=1,2,\cdots,7 \right)$ in Eq.~(\ref{eq:I}) are given by
\begin{eqnarray}
        k^2 I^{(2)}_1\left(u,v,x\to \infty \right)& =& 
        \frac{27 (u^2 + v^2 - 3)}{u^3 v^3 x}
        \bigg(
            \sin x 
            \big(
                -4 u v + (u^2 + v^2 - 3) 
                \ln \left| \frac{3 - (u + v)^2}{3 - (u - v)^2} \right|
            \big) \nonumber\\
        &-& \pi (u^2 + v^2 - 3) \Theta(v + u - \sqrt{3}) \cos x
        \bigg) \ , \label{eq:11I}\\[6pt]
       k^2 I^{(2)}_2\left(u,v,x \to \infty\right)&=& 
        \frac{\sqrt{3} \left(u^2 - 3 (1-v)^2\right)}{16 u^3 v x}
        \Bigg[
            \sin x 
            \bigg(
                \left(u^2 - 3 (1+v)^2\right) 
                \ln \left|
                    \frac{\left(u - \sqrt{3} v\right)^2 - 3}{\left(u + \sqrt{3} v\right)^2 - 3}
                \right| \nonumber\\
         &-&\frac{4 u v \left(u^2 - 9 v^2 + 9\right)}{\sqrt{3} \left(u^2 - 3 (1-v)^2\right)}
            \bigg) - \pi \left(u^2 - 3 (1+v)^2\right) \nonumber\\
        &\times&~\Theta (u +  \sqrt{3}v - \sqrt{3}) \cos x 
        \Bigg]\ , \label{eq:22I}\\[6pt]
        k^2 I^{(2)}_3\left(u,v,x \right)&=&
        \frac{\sin x}{4 x} 
        - \frac{\sin(u x) \sin(v x)}{4 u v x^2}\ , \label{eq:33I}\\[8pt]
        k^2 I^{(2)}_i\left(u,v,x \right)&=&  -\frac{1}{8uvx}\Bigg[\sin x \bigg( \text{Ci}\big((1 - u + v)x\big)+\text{Ci}\big( (1 + u - v)x \big) \nonumber \\
        &&-\, \text{Ci}\big(\left|1 - u - v\right|x\big)-\text{Ci}\big((1 + u + v)x \big) +\ln\left|\frac{1 - (u + v)^2}{1 - (u - v)^2}\right| \bigg) \nonumber \\
        &&- \, \cos x \bigg( \text{Si}\big((1 - u + v)x\big )-\text{Si}\big((1 - u -v)x \big ) \nonumber \\
        && +\, \text{Si}\big((1 + u - v)x \big)-\text{Si} \big( (1 + u + v)x \big) \bigg)
        \Bigg]\  , \ \quad (i=4,5,6,7) \ , \label{eq:ijI}
\end{eqnarray}
where we have used the following approximations: $\lim_{x\to\pm \infty} \mathrm{Si}(x)=\pm \pi/2$ and $\lim_{x\to \infty} \mathrm{Ci}(x)=0$ in Eq.~(\ref{eq:11I}) and Eq.~(\ref{eq:22I}). $\Theta(x)$ represents the Heaviside theta function. Eq.~(\ref{eq:33I}) and Eq.~(\ref{eq:ijI}) provide the analytical expressions for the kernel functions corresponding to the first-order tensor source term. Similar to Eq.~(\ref{eq:11I}) and Eq.~(\ref{eq:22I}), when calculating the energy density spectrum of gravitational waves, we can use the oscillating average relations $\lim_{x\to \infty}\sin^2x=1/2$, $\lim_{x\to \infty}\cos^2x=1/2$, and $\lim_{x\to \infty}\cos{x}\sin{x}=0$ to derive the approximate expression for the kernel function when $x\to \infty$ \cite{Yuan:2021qgz}. In this section, we derived the equations of motion of second-order \acp{TSIGW} using higher-order cosmological perturbation theory and solved them in momentum space. It is worth noting that we can also directly obtain the specific expressions for second-order \acp{TSIGW} in Eq.~(\ref{eq:h1})--Eq.~(\ref{eq:h7}) using the vertex rules of higher-order cosmological perturbations, without deriving or solving the second-order cosmological perturbation equations. Detailed discussion can be found in Ref.~\cite{Zhou:2024ncc}. Moreover, the specific expressions of the second-order kernel functions $I^{(2)}_i(u,v,x)$ given above apply to \acp{TSIGW} during the \ac{RD} era. These results can be extended to the study of \acp{TSIGW} in other dominant eras, such as the early matter-dominated era \cite{Assadullahi:2009nf,Alabidi:2013lya,Domenech:2020ssp,Papanikolaou:2020qtd}.

\section{Energy density spectra of TSIGWs}\label{sec:3.0}
\begin{figure*}[htbp]
    \captionsetup{
      justification=raggedright,
      singlelinecheck=true
    }
    \centering
	\subfloat[$\langle \zeta_{\mathbf{k}-\mathbf{p}}\zeta_{\mathbf{k}'-\mathbf{p}'} \rangle \langle \zeta_{\mathbf{p}}\zeta_{\mathbf{p}'} \rangle$ \label{fig:Feyn22a}]{\includegraphics[width=.3\columnwidth]{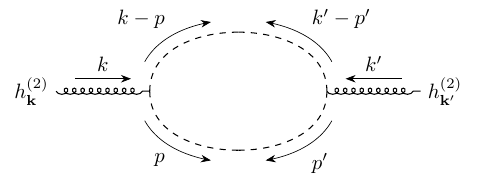}}
        \subfloat[$\langle \zeta_{\mathbf{k}-\mathbf{p}}\zeta_{\mathbf{p}'} \rangle \langle \zeta_{\mathbf{p}}\zeta_{\mathbf{k}'-\mathbf{p}'} \rangle$\label{fig:Feyn22b}]{\includegraphics[width=.3\columnwidth]{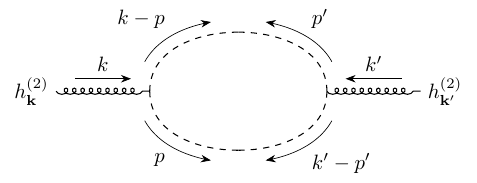}}
        \subfloat[$\langle \zeta_{\mathbf{k}-\mathbf{p}}\zeta_{\mathbf{p}} \rangle \langle \zeta_{\mathbf{k}'-\mathbf{p}'} \zeta_{\mathbf{p}'} \rangle$\label{fig:Feyn22c}]{\includegraphics[width=.35\columnwidth]{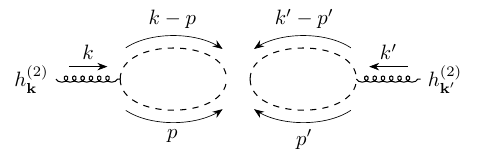}} \\

        \subfloat[$\langle \mathbf{h}_{\mathbf{k}-\mathbf{p}}\mathbf{h}_{\mathbf{k}'-\mathbf{p}'} \rangle \langle \mathbf{h}_{\mathbf{p}}\mathbf{h}_{\mathbf{p}'} \rangle$\label{fig:Feyn22d}]{\includegraphics[width=.3\columnwidth]{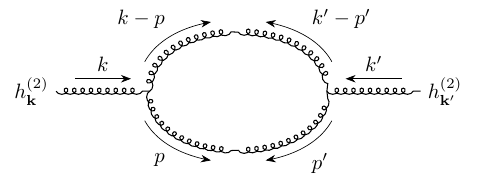}}
        \subfloat[$\langle \mathbf{h}_{\mathbf{k}-\mathbf{p}}\mathbf{h}_{\mathbf{p}'} \rangle \langle \mathbf{h}_{\mathbf{p}}\mathbf{h}_{\mathbf{k}'-\mathbf{p}'} \rangle$\label{fig:Feyn22e}]{\includegraphics[width=.3\columnwidth]{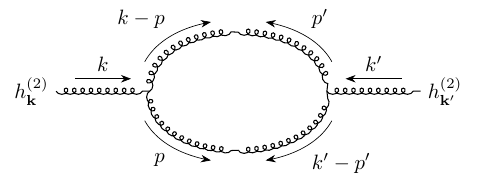}}
        \subfloat[$\langle \mathbf{h}_{\mathbf{k}-\mathbf{p}}\mathbf{h}_{\mathbf{p}} \rangle \langle \mathbf{h}_{\mathbf{k}'-\mathbf{p}'} \mathbf{h}_{\mathbf{p}'} \rangle$\label{fig:Feyn22f}]{\includegraphics[width=.35\columnwidth]{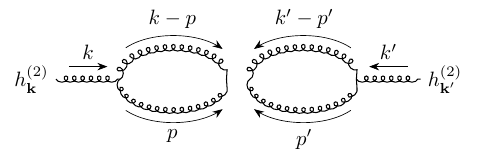}} \\
        \subfloat[$\langle \zeta_{\mathbf{k}-\mathbf{p}}\zeta_{\mathbf{k}'-\mathbf{p}‘} \rangle \langle \mathbf{h}_{\mathbf{p}} \mathbf{h}_{\mathbf{p}'} \rangle$\label{fig:Feyn22g}]{\includegraphics[width=.3\columnwidth]{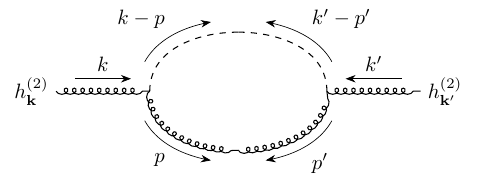}} 
\caption{\label{fig:FeynDiag22} The diagrams correspond to the Wick’s expansions of four-point correlation functions of primordial perturbations. The dashed and spring-like lines in the figure represent scalar and tensor perturbations, respectively. }
\end{figure*}

The total energy density fraction of \acp{GW} up to second order can be written as \cite{Maggiore:1999vm}
\begin{equation}\label{eq:Omega}
	\begin{aligned}
		\Omega_{\mathrm{GW}}^{\mathrm{tot}}(\eta, k)&=\frac{\rho^{(1)}_{\mathrm{GW}}(\eta, k)+\frac{1}{4}\rho^{(2)}_{\mathrm{GW}}(\eta, k)}{\rho_{\mathrm{tot}}(\eta)} \\
		&=\frac{1}{6}\left(\frac{k}{a(\eta) H(\eta)}\right)^{2} \left( \mathcal{P}^{(1)}_{h}(\eta, k)+\frac{1}{4}\mathcal{P}^{(2)}_{h}(\eta, k) \right) \ ,
	\end{aligned}
\end{equation}
where  
\begin{equation}\label{eq:P(1)}
    \mathcal{P}^{(1)}_{h}(\eta, k)=\mathcal{P}^{(1)}_h\left( k \right) \left( T_h\left( x \right) \right)^2 
\end{equation}
is the primordial power spectrum of first-order \acp{GW} $h^{(1)}_{ij}$. $\mathcal{P}^{(1)}_h(k)$ in Eq.~(\ref{eq:P(1)}) represents the power spectrum of \acp{PGW}. In Eq.~(\ref{eq:Omega}),  $\mathcal{P}^{(2)}_{h}(\eta, k)$ represents the power spectrum of second-order \acp{TSIGW}. The power spectra of $n$-th order \acp{GW} $\mathcal{P}_{h}^{(n)}( \mathbf{k},\eta)$ in Eq.~(\ref{eq:Omega}) are defined as
\begin{equation}\label{eq:Ph}
  \left\langle h^{\lambda,(n)}( \mathbf{k},\eta) h^{\lambda^{\prime},(n)}\left(\mathbf{k}^{\prime},\eta\right)\right\rangle=\delta^{\lambda \lambda^{\prime}} \delta\left(\mathbf{k}+\mathbf{k}^{\prime}\right) \frac{2 \pi^2}{k^3} \mathcal{P}_{h}^{(n)}(k,\eta) \ .  
\end{equation}
As shown in Eq.~(\ref{eq:Ph}), the power spectrum of \acp{GW} can be calculated in terms of the  corresponding two-point function. By substituting Eq.~(\ref{eq:h1})--Eq.~(\ref{eq:h7}) into Eq.~(\ref{eq:Ph}), we obtain the explicit expression of the power spectra of second order \acp{TSIGW} \cite{Chang:2022vlv}
\begin{equation}\label{eq:Psum}
\mathcal{P}^{(2)}_h=\mathcal{P}^{(2),11}_h+\mathcal{P}^{(2),22}_h+\sum_{i,j=3}^{7}\mathcal{P}^{(2),ij}_h \ ,
\end{equation}
where
\begin{eqnarray}\label{eq:P11} 
		\mathcal{P}^{(2),11}_h&=&\frac{1}{4}\int_{0}^{\infty} \mathrm{d}v\int_{|1-v|}^{|1+v|}\mathrm{d}u \left( \frac{ 4v^2-\left( 1+v^2-u^2 \right)^2}{4uv} \right)^2 \left( k^2 I^{(2)}_1\left( u,v,x \right)  \right)^2 \nonumber\\
  & &\times~ \frac{4}{9}~ \mathcal{P}^{(1)}_{\zeta}\left( ku\right) ~\frac{4}{9}~ \mathcal{P}^{(1)}_{\zeta}\left( vk \right)\ ,
  \end{eqnarray}
  \begin{eqnarray}\label{eq:P22}
		\mathcal{P}^{(2),22}_h&=&\frac{1}{4}\int_{0}^{\infty} \mathrm{d}v\int_{|1-v|}^{|1+v|}\mathrm{d}u ~ \frac{1}{64(uv)^2} \left(\frac{16 \left(-u^2+v^2+1\right)^2}{v^2}+\left(\frac{\left(-u^2+v^2+1\right)^2}{v^2}+4\right)^2\right) \nonumber\\
	  	& &\times~ \frac{4}{9}~  \left( k^2 I^{(2)}_2\left( u,v,x \right)  \right)^2 \mathcal{P}^{(1)}_{\zeta}\left( ku\right) \mathcal{P}_{h}^{(1)}\left( vk \right) \ ,
\end{eqnarray}
\begin{eqnarray}\label{eq:Pij} 
	\mathcal{P}^{(2),ij}_h&=&\frac{1}{4}\int_{0}^{\infty} \mathrm{d}v\int_{|1-v|}^{|1+v|}\mathrm{d}u~\frac{\mathbb{P}^{ij}\left( u,v \right)}{\left(uv\right)^2}  \left( k^2 I^{(2)}_i\left( u,v,x \right)  k^2I^{(2)}_j\left( u,v,x \right)  \right) \nonumber\\
	  	& &\times~ \mathcal{P}_{h}^{(1)}\left( ku\right) \mathcal{P}_{h}^{(1)}\left( vk \right) \ , \  \left(i,j=3\sim 7\right) \ .
\end{eqnarray}
In Eq.~(\ref{eq:P11})--Eq.~(\ref{eq:Pij}), $\mathcal{P}^{(2),11}_h$, $\mathcal{P}^{(2),22}_h$, and $\mathcal{P}^{(2),ij}_h$ represent the second-order power spectra from scalar-scalar, scalar-tensor, and tensor-tensor coupling source terms, respectively. These contributions depend on the amplitudes and shapes of the primordial power spectra $\mathcal{P}^{(1)}_{\zeta}(k)$ and $\mathcal{P}^{(1)}_{h}(k)$. As shown in Fig.~\ref{fig:FeynDiag22}, the power spectra of second-order \acp{TSIGW} correspond to the one-loop diagrams in higher-order cosmological perturbation theory \cite{Zhou:2024ncc}. More precisely, Fig.~(\ref{fig:Feyn22a}) $\sim$ Fig.~(\ref{fig:Feyn22c}) correspond to the power spectrum of second-order \acp{SIGW} in Eq.~(\ref{eq:P11}), Fig.~(\ref{fig:Feyn22d}) $\sim$ Fig.~(\ref{fig:Feyn22f}) correspond to the power spectrum of second-order \acp{GW} induced by primordial tensor perturbations in Eq.~(\ref{eq:Pij}), and Fig.~(\ref{fig:Feyn22g}) corresponds to the power spectrum of second-order \acp{GW} induced by tensor-scalar source terms in Eq.~(\ref{eq:P22}).  Since the integration over the azimuth angle of momentum $\mathbf{p}$ is zero, Fig.~(\ref{fig:Feyn22c}) and Fig.~(\ref{fig:Feyn22f}) do not contribute to the power spectrum of \acp{TSIGW}. The momentum  polynomials $\mathbb{P}^{ij}\left( u,v \right)$ in Eq.~(\ref{eq:Pij}) are given in Table.~\ref{ta:1}. In the calculation of the power spectra $\mathcal{P}^{(2),ij}_h$, we encounter the four-point correlation function of \acp{PGW}  
\begin{equation}\label{eq:A2}
	\begin{aligned}
		\langle \mathbf{h}_{\mathbf{k}-\mathbf{p}}^{\lambda_1}\mathbf{h}_{\mathbf{p}}^{\lambda_2} \mathbf{h}_{\mathbf{k}'-\mathbf{p}'}^{\lambda_1'}\mathbf{h}_{\mathbf{p}'}^{\lambda_2'} \rangle&=\langle \mathbf{h}_{\mathbf{k}-\mathbf{p}}^{\lambda_1} \mathbf{h}_{\mathbf{k}'-\mathbf{p}'}^{\lambda_1'}  \rangle\langle \mathbf{h}_{\mathbf{p}}^{\lambda_2}\mathbf{h}_{\mathbf{p}'}^{\lambda_2'} \rangle+\langle \mathbf{h}_{\mathbf{k}-\mathbf{p}}^{\lambda_1}\mathbf{h}_{\mathbf{p}'}^{\lambda_2'} \rangle \langle \mathbf{h}_{\mathbf{p}}^{\lambda_2} \mathbf{h}_{\mathbf{k}'-\mathbf{p}'}^{\lambda_1'} \rangle \\
		&=\delta\left(\mathbf{k}+\mathbf{k}'\right) \frac{(2\pi^2)^2}{p^3|\mathbf{k}-\mathbf{p}|^3} \left(\delta^{\lambda_1\lambda'_1}\delta^{\lambda_2\lambda'_2}\delta\left(\mathbf{p}+\mathbf{p}'\right)+\delta^{\lambda_1\lambda'_2}\delta^{\lambda_2\lambda'_1}\delta\left(\mathbf{k}'-\mathbf{p}'+\mathbf{p}\right) \right)\\
        &\times ~\mathcal{P}_{h}(|\mathbf{k}-\mathbf{p}|)\mathcal{P}_{h}(|\mathbf{p}|) \ .
	\end{aligned}
\end{equation}
By integrating over the momentum $\mathbf{p}'$, we derive the substitution relation: $\mathbf{p}'\to -\mathbf{p}$ and  $\mathbf{p}'\to \mathbf{p}-\mathbf{k}$. Then, we can express the momentum polynomial $\mathbb{P}^{ij}$ in Table.~\ref{ta:1} as a function of $u=|\mathbf{k}-\mathbf{p}|/|\mathbf{k}|$ and $v=|\mathbf{p}|/|\mathbf{k}|$.
\begin{table}[h!]
\centering
\begin{tabular}{|c|c|c|c|c|}
\hline 
\hline
$k^4\mathbb{P}^{i j}$ & $i=3$   \\
\hline
$j=3$ & $\delta^{\lambda \lambda'} \left(k^2 \varepsilon^{\lambda,lm}\left(\mathbf{k}\right) \varepsilon^{\lambda_1,b}_{l}\left(\mathbf{k}-\mathbf{p}\right)\varepsilon^{\lambda_2}_{bm}\left(\mathbf{p}\right) \right)\left(k'^2 \varepsilon^{\lambda',lm}\left(\mathbf{k}'\right) \varepsilon^{\lambda_1',b}_{l}\left(\mathbf{k}'-\mathbf{p}'\right)\varepsilon^{\lambda_2'}_{bm}\left(\mathbf{p}'\right) \right)$   \\
\hline
$j=4$ & $\delta^{\lambda \lambda'}\left(k^2 \varepsilon^{\lambda,lm}\left(\mathbf{k}\right) \varepsilon^{\lambda_1,b}_{l}\left(\mathbf{k}-\mathbf{p}\right)\varepsilon^{\lambda_2}_{bm}\left(\mathbf{p}\right) \right)\left( \varepsilon^{\lambda',lm}\left(\mathbf{k}'\right) \varepsilon^{\lambda_1',bc}\left( \mathbf{k}'-\mathbf{p}' \right)\left(  2\varepsilon^{\lambda_2'}_{mb}\left( \mathbf{p}' \right)p'_cp'_l\right)  \right)$  \\
\hline
$j=5$ & $\delta^{\lambda \lambda'}\left(k^2\varepsilon^{\lambda,lm}\left(\mathbf{k}\right) \varepsilon^{\lambda_1,b}_{l}\left(\mathbf{k}-\mathbf{p}\right)\varepsilon^{\lambda_2}_{bm}\left(\mathbf{p}\right) \right)\left( \varepsilon^{\lambda',lm}\left(\mathbf{k}'\right) \varepsilon^{\lambda_1'}_{mc}\left(\mathbf{k}'-\mathbf{p}'\right)\varepsilon^{\lambda_2'}_{lb}\left(\mathbf{p}'\right)\left( k'-p' \right)^bp'^c  \right)$  \\
\hline
$j=6$ & $\delta^{\lambda \lambda'}\left( k^2\varepsilon^{\lambda,lm}\left(\mathbf{k}\right) \varepsilon^{\lambda_1,b}_{l}\left(\mathbf{k}-\mathbf{p}\right)\varepsilon^{\lambda_2}_{bm}\left(\mathbf{p}\right) \right)\left(  \varepsilon^{\lambda',lm}\left(\mathbf{k}'\right) \varepsilon^{\lambda_1'}_{bc}\left(\mathbf{k}'-\mathbf{p}'\right)\varepsilon^{\lambda_2'}_{lm}\left(\mathbf{p}'\right)p'^bp'^c \right)$  \\
\hline
$j=7$ & $\delta^{\lambda \lambda'}\left( k^2\varepsilon^{\lambda,lm}\left(\mathbf{k}\right) \varepsilon^{\lambda_1,b}_{l}\left(\mathbf{k}-\mathbf{p}\right)\varepsilon^{\lambda_2}_{bm}\left(\mathbf{p}\right) \right)\left( \varepsilon^{\lambda',lm}\left(\mathbf{k}'\right) \varepsilon^{\lambda_1',bc}\left(\mathbf{k}'-\mathbf{p}'\right)\varepsilon^{\lambda_2'}_{bc}\left(\mathbf{p}'\right)p'_lp'_m \right)$  \\
\hline
\hline
\hline
$k^4\mathbb{P}^{i j}$ & $i=4$   \\
\hline
$j=4$ & $\delta^{\lambda \lambda'}\left( \varepsilon^{\lambda,lm}\left(\mathbf{k}\right) \varepsilon^{\lambda_1,bc}\left( \mathbf{k}-\mathbf{p} \right)\left(  2\varepsilon^{\lambda_2}_{mb}\left( \mathbf{p} \right)p_cp_l\right)  \right)\left( \varepsilon^{\lambda',lm}\left(\mathbf{k}'\right) \varepsilon^{\lambda_1',bc}\left( \mathbf{k}'-\mathbf{p}' \right)\left(  2\varepsilon^{\lambda_2'}_{mb}\left( \mathbf{p}' \right)p'_cp'_l\right) \right)$   \\
\hline
$j=5$ & $\delta^{\lambda \lambda'}\left( \varepsilon^{\lambda,lm}\left(\mathbf{k}\right) \varepsilon^{\lambda_1,bc}\left( \mathbf{k}-\mathbf{p} \right)\left(  2\varepsilon^{\lambda_2}_{mb}\left( \mathbf{p} \right)p_cp_l\right)  \right)\left(  \varepsilon^{\lambda',lm}\left(\mathbf{k}'\right) \varepsilon^{\lambda_1'}_{mc}\left(\mathbf{k}'-\mathbf{p}'\right)\varepsilon^{\lambda_2'}_{lb}\left(\mathbf{p}'\right)\left( k'-p' \right)^bp'^c \right)$  \\
\hline
$j=6$ & $\delta^{\lambda \lambda'}\left( \varepsilon^{\lambda,lm}\left(\mathbf{k}\right) \varepsilon^{\lambda_1,bc}\left( \mathbf{k}-\mathbf{p} \right)\left(  2\varepsilon^{\lambda_2}_{mb}\left( \mathbf{p} \right)p_cp_l\right)  \right)\left( \varepsilon^{\lambda',lm}\left(\mathbf{k}'\right) \varepsilon^{\lambda_1'}_{bc}\left(\mathbf{k}'-\mathbf{p}'\right)\varepsilon^{\lambda_2'}_{lm}\left(\mathbf{p}'\right)p'^bp'^c  \right)$  \\
\hline
$j=7$ & $\delta^{\lambda \lambda'}\left( \varepsilon^{\lambda,lm}\left(\mathbf{k}\right) \varepsilon^{\lambda_1,bc}\left( \mathbf{k}-\mathbf{p} \right)\left(  2\varepsilon^{\lambda_2}_{mb}\left( \mathbf{p} \right)p_cp_l\right)  \right)\left( \varepsilon^{\lambda',lm}\left(\mathbf{k}'\right) \varepsilon^{\lambda_1',bc}\left(\mathbf{k}'-\mathbf{p}'\right)\varepsilon^{\lambda_2'}_{bc}\left(\mathbf{p}'\right)p'_lp'_m \right)$  \\
\hline
\hline
\hline
$k^4\mathbb{P}^{i j}$ & $i=5$   \\
\hline
$j=5$ & $\delta^{\lambda \lambda'}\left( \varepsilon^{\lambda,lm}\left(\mathbf{k}\right) \varepsilon^{\lambda_1}_{mc}\left(\mathbf{k}-\mathbf{p}\right)\varepsilon^{\lambda_2}_{lb}\left(\mathbf{p}\right)\left( k-p \right)^bp^c  \right)\left(  \varepsilon^{\lambda',lm}\left(\mathbf{k}'\right) \varepsilon^{\lambda_1'}_{mc}\left(\mathbf{k}'-\mathbf{p}'\right)\varepsilon^{\lambda_2'}_{lb}\left(\mathbf{p}'\right)\left( k'-p' \right)^bp'^c \right)$   \\
\hline
$j=6$ & $\delta^{\lambda \lambda'}\left( \varepsilon^{\lambda,lm}\left(\mathbf{k}\right) \varepsilon^{\lambda_1}_{mc}\left(\mathbf{k}-\mathbf{p}\right)\varepsilon^{\lambda_2}_{lb}\left(\mathbf{p}\right)\left( k-p \right)^bp^c   \right)\left( \varepsilon^{\lambda',lm}\left(\mathbf{k}'\right) \varepsilon^{\lambda_1'}_{bc}\left(\mathbf{k}'-\mathbf{p}'\right)\varepsilon^{\lambda_2'}_{lm}\left(\mathbf{p}'\right)p'^bp'^c \right)$  \\
\hline
$j=7$ & $\delta^{\lambda \lambda'}\left( \varepsilon^{\lambda,lm}\left(\mathbf{k}\right) \varepsilon^{\lambda_1}_{mc}\left(\mathbf{k}-\mathbf{p}\right)\varepsilon^{\lambda_2}_{lb}\left(\mathbf{p}\right)\left( k-p \right)^bp^c  \right)\left( \varepsilon^{\lambda',lm}\left(\mathbf{k}'\right) \varepsilon^{\lambda_1',bc}\left(\mathbf{k}'-\mathbf{p}'\right)\varepsilon^{\lambda_2'}_{bc}\left(\mathbf{p}'\right)p'_lp'_m  \right)$  \\
\hline
\hline
\hline
$k^4\mathbb{P}^{i j}$ & $i=6$   \\
\hline
$j=6$ & $\delta^{\lambda \lambda'}\left( \varepsilon^{\lambda,lm}\left(\mathbf{k}\right) \varepsilon^{\lambda_1}_{bc}\left(\mathbf{k}-\mathbf{p}\right)\varepsilon^{\lambda_2}_{lm}\left(\mathbf{p}\right)p^bp^c \right)\left( \varepsilon^{\lambda',lm}\left(\mathbf{k}'\right) \varepsilon^{\lambda_1'}_{bc}\left(\mathbf{k}'-\mathbf{p}'\right)\varepsilon^{\lambda_2'}_{lm}\left(\mathbf{p}'\right)p'^bp'^c \right)$   \\
\hline
$j=7$ & $\delta^{\lambda \lambda'}\left( \varepsilon^{\lambda,lm}\left(\mathbf{k}\right) \varepsilon^{\lambda_1}_{bc}\left(\mathbf{k}-\mathbf{p}\right)\varepsilon^{\lambda_2}_{lm}\left(\mathbf{p}\right)p^bp^c \right)\left( \varepsilon^{\lambda',lm}\left(\mathbf{k}'\right) \varepsilon^{\lambda_1',bc}\left(\mathbf{k}'-\mathbf{p}'\right)\varepsilon^{\lambda_2'}_{bc}\left(\mathbf{p}'\right)p'_lp'_m \right)$  \\
\hline
\hline
\hline
$k^4\mathbb{P}^{i j}$ & $i=7$   \\
\hline
$j=7$ & $\delta^{\lambda \lambda'}\left( \varepsilon^{\lambda,lm}\left(\mathbf{k}\right) \varepsilon^{\lambda_1,bc}\left(\mathbf{k}-\mathbf{p}\right)\varepsilon^{\lambda_2}_{bc}\left(\mathbf{p}\right)p_lp_m  \right)\left( \varepsilon^{\lambda',lm}\left(\mathbf{k}'\right) \varepsilon^{\lambda_1',bc}\left(\mathbf{k}'-\mathbf{p}'\right)\varepsilon^{\lambda_2'}_{bc}\left(\mathbf{p}'\right)p'_lp'_m \right)$  \\
\hline
\end{tabular}
\caption{The explicit expressions of momentum polynomials $\mathbb{P}^{i j}(\mathbf{k},\mathbf{p},\mathbf{k}',\mathbf{p}')$.  }
\label{ta:1}
\end{table}

\begin{figure}[!ht]
\centering
\includegraphics[width=0.7\linewidth]{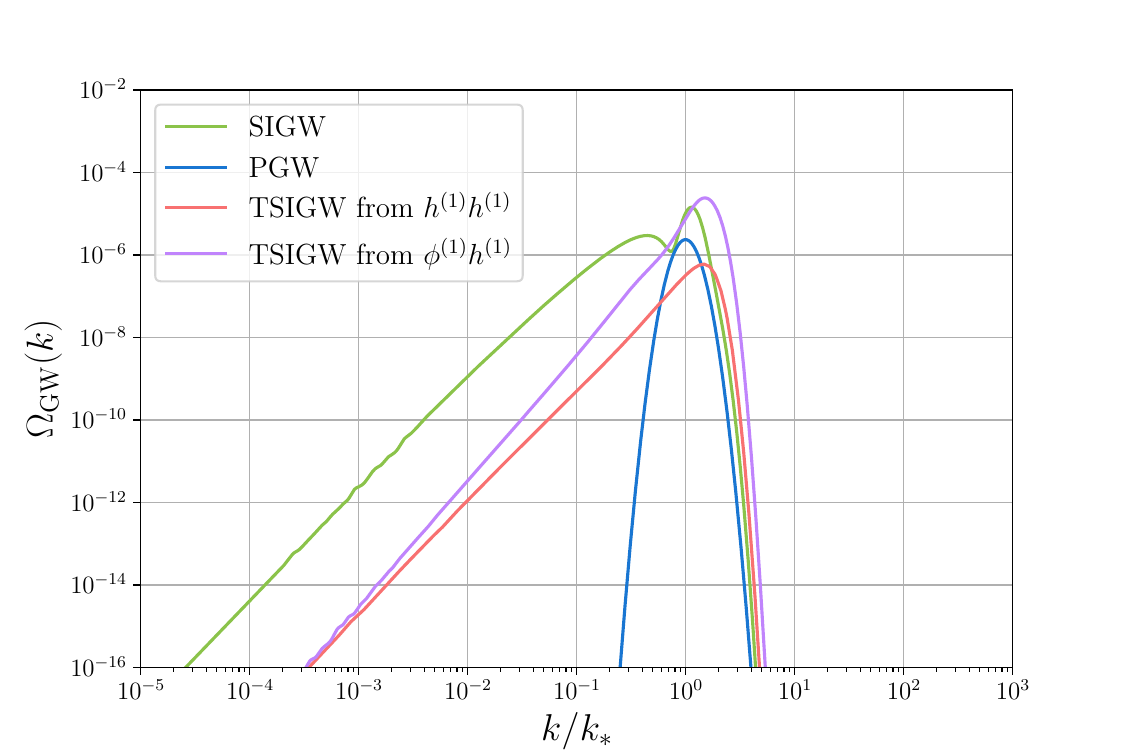}
\caption{The energy spectra contributions from \acp{SIGW}, \acp{PGW}, tensor-tensor sources, and tensor-scalar sources are depicted by the green, blue, red, and purple lines, respectively. Parameters are set to $A_{\zeta}=A_h=1$ and $\sigma_*=0.2$. } \label{fig:spectrum}
\end{figure}

In this paper, we consider the primordial power spectra $\mathcal{P}^{(1)}_{\zeta}(k)$ and $\mathcal{P}^{(1)}_{h}(k)$ with log-normal peaks at the same location $k_*$ on small scales
\begin{eqnarray}\label{eq:Pn1}
	\mathcal{P}^{(1)}_{\zeta}(k)=\frac{A_{\zeta}}{\sqrt{2 \pi \sigma_{*}^2}} \exp \left(-\frac{\ln^2\left(k / k_*\right)}{2 \sigma_{*}^2}\right) \ ,  \nonumber\\ \mathcal{P}^{(1)}_{h}(k)=\frac{A_{h}}{\sqrt{2 \pi \sigma_{*}^2}} \exp \left(-\frac{\ln^2 \left(k / k_*\right)}{2 \sigma_{*}^2}\right) \ ,
\end{eqnarray}
where $k_*$ is the wave number at which the power spectrum has a log-normal peak. $A_{\zeta}$ and $A_{h}$ are amplitudes of primordial curvature and tensor power spectra respectively. $\sigma_{*}$ is the parameter describing the widths of primordial power spectra. In the limit of $\sigma_{*}\to 0$, $\mathcal{P}^{(1)}_{\zeta}(k)$ and $\mathcal{P}^{(1)}_{h}(k)$ reduce to the monochromatic power spectra. Taking into account the thermal history of the universe, the current total energy density spectrum $\bar{\Omega}^{\mathrm{tot}}_{\mathrm{GW,0}}(k)$ is given by
\begin{equation}
    \bar{\Omega}^{\mathrm{tot}}_{\mathrm{GW,0}}(k) = \Omega_{\mathrm{rad},0}\left(\frac{g_{*,\rho,e}}{g_{*,\rho,0}}\right)\left(\frac{g_{*,s,0}}{g_{*,s,e}}\right)^{4/3}\bar{\Omega}^{\mathrm{tot}}_{\mathrm{GW}}(\eta,k) \ ,
\end{equation}
where $\Omega_{\mathrm{rad},0}$ ($ =4.2\times 10^{-5}h^{-2}$) is the energy density fraction of radiations today. Fig.~\ref{fig:spectrum} presents a comparison between the energy density spectrum of second-order \acp{SIGW} and the energy density spectra described in Eq.~(\ref{eq:Omega}).

\section{Detection of TSIGWs}\label{sec:4.0}
In this section, we analyze how cosmological observations across various scales constrain second-order \acp{TSIGW}. Specifically, we consider the \acp{SGWB} dominated by \acp{PGW}$+$\acp{TSIGW}. By incorporating the energy density spectrum of \acp{TSIGW} studied in Sec.~\ref{sec:3.0}, we perform a Bayesian analysis to current \ac{PTA} observations to determine or limit the parameter space of the primordial power spectrum. Since the tensor-scalar and tensor-tensor source terms significantly influence the total energy density spectrum of \acp{GW} at high frequencies, we thoroughly investigate the impact of small-scale \acp{PGW}$+$\acp{TSIGW} on \ac{SNR} of \ac{LISA}. Furthermore, \acp{PGW} and \acp{TSIGW} can serve as an extra radiation component, influencing the  \ac{CMB} and \ac{BAO} observations. Based on current \ac{CMB} and \ac{BAO} observational data, we can constrain the parameter space of the small-scale primordial power spectrum.

\subsection{PTA observations}\label{sec:4.1}
\begin{figure}[htbp]
    \centering
    \includegraphics[width=.9\columnwidth]{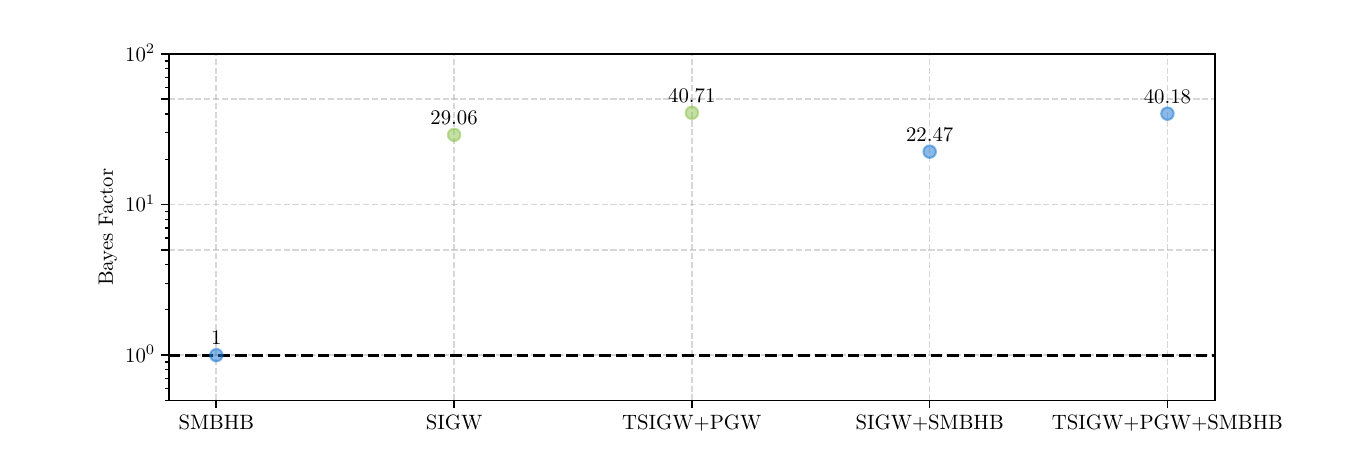}
\caption{\label{fig:bayes} The Bayes factors between different models. The vertical axis represents the Bayes factor of different models relative to \ac{SMBHB}, and the horizontal axis represents the different models. The green dots are for models without \ac{SMBHB} and the blue dots are for models in combination with the \ac{SMBHB} signal. }
\end{figure}

To constrain the parameter space of primordial power spectra from \ac{PTA} observations, we employ the \ac{KDE} representations of the free spectra to construct the likelihood function \cite{Mitridate:2023oar,Lamb:2023jls,Moore:2021ibq}
\begin{equation} \label{eq:likelihood}
    \ln \mathcal{L}(d|\theta) = \sum_{i=1}^{N_f} p(\Phi_i,\theta)\ .
\end{equation}
Here $p(\Phi_i,\theta)$ represents the probability of $\Phi_i$ given the parameter $\theta$, and $\Phi_i = \Phi(f_i)$ denotes the time delay
\begin{equation} \label{eq:timedelay}
    \Phi(f) = \sqrt{\frac{H_0^2 \Omega_{\mathrm{GW}}(f)}{8\pi^2 f^5 T_{\mathrm{obs}}}} \ ,
\end{equation}
where $H_0=h\times 100 \mathrm{km/s/Mpc}$ is the present-day value of the Hubble constant. We directly use the \acp{KDE} representation of the first 14 frequency HD-correlated free spectrum in NANOGrav 15-year dataset \cite{Nanograv:KDE}. Bayesian analysis is performed by \textsc{bilby} \cite{bilby_paper} with its built-in \textsc{dynesty} nested sampler \cite{Speagle:2019ivv,dynesty_software}.
The spectrum of \acp{SMBHB} is modeled as power law \cite{Mitridate:2023oar,NANOGrav:2023hvm}
\begin{equation} \label{eq:SMBHB}
    \Omega_{\mathrm{GW}}(f) = \frac{2\pi^2 A_{\mathrm{BHB}}^2}{3H_0^2 h^2} (\frac{f}{\mathrm{year}^{-1}})^{5-\gamma_{\mathrm{BHB}}}\mathrm{year}^{-2} \ .
\end{equation}
The prior distribution for $(\log_{10}A_{\mathrm{BHB}}, \gamma_{\mathrm{BHB}})$ follows a multivariate normal distribution \cite{NANOGrav:2023hvm}, whose mean and covariance matrix are given by
\begin{equation} \label{eq:prior_SMBHB}
    \boldsymbol{\mu}_{\mathrm{BHB}} =\begin{pmatrix} -15.6
 \\ 4.7 \end{pmatrix} , 
\boldsymbol{\sigma}_{\mathrm{BHB}}=0.1\times \begin{pmatrix}
2.8  & -0.026\\
-0.026  & 2.8
\end{pmatrix} \ .
\end{equation} 
We employ the Bayes factor to compare different models.
The Bayes factor is defined as $B_{i,j} = \frac{Z_i}{Z_j}$, where $Z_i$ represents the evidence of model $H_i$. Fig. \ref{fig:bayes} illustrates Bayes factors for comparisons between various models and the \ac{SMBHB} model. As shown in Fig.~\ref{fig:bayes}, the Bayes factor corresponding to \acp{PGW}$+$\acp{TSIGW}$+$\ac{SMBHB} is the largest. Therefore, \acp{PGW}+\acp{TSIGW} are more likely to dominate the \acp{SGWB} observed by current \ac{PTA} experiments than \ac{SMBHB}.

\begin{figure*}[htbp]
    \captionsetup{
      justification=raggedright,
      singlelinecheck=true
    }
    \centering
	\subfloat[\acp{SIGW} and \acp{SIGW}$+$
\ac{SMBHB}.\label{fig:corner_SIGW}]  {\includegraphics[width=.49\columnwidth]{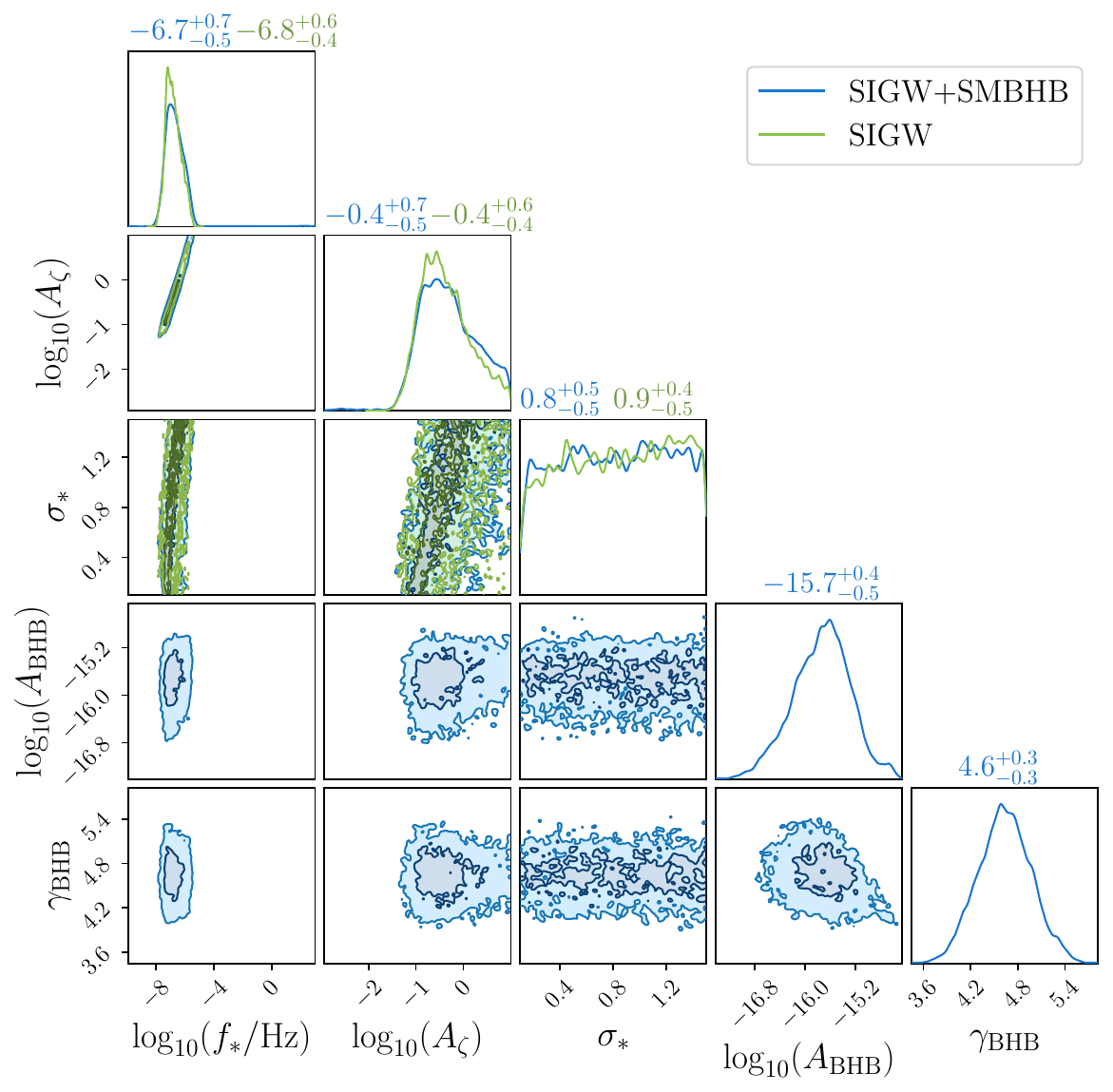}}
    \hspace{0.05cm}
        \subfloat[\acp{PGW}$+$\acp{TSIGW} and \acp{PGW}$+$\acp{TSIGW}$+$
\ac{SMBHB}.\label{fig:corner_TSIGW}]{\includegraphics[width=.49\columnwidth]{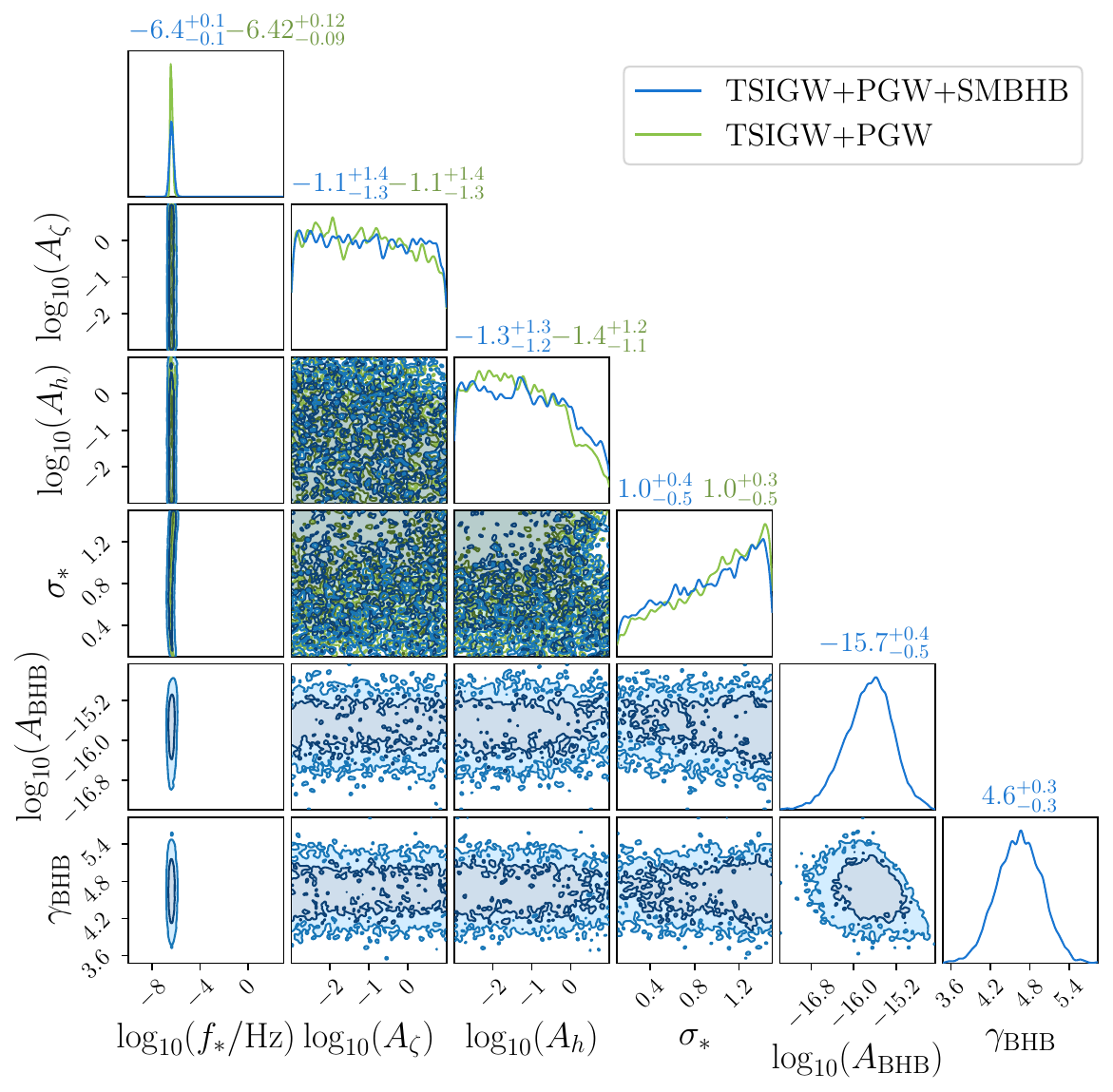}} 
\caption{The corner plot of the posterior distributions. The contours in the off-diagonal panels denote the $68\% $ and $95 \%$ credible intervals of the 2D posteriors. The numbers above the figures represent the median values and $1$-$\sigma$ ranges of the parameters.  \textbf{Left panel: }The blue and green solid curves correspond to the \ac{SIGW} energy spectrum with or without \ac{SMBHB}. \textbf{Right panel: } The blue and green solid curves correspond to the \acp{PGW}$+$\ac{TSIGW} energy spectrum with or without \ac{SMBHB}.  }
\end{figure*}

\begin{figure}[!ht]
\centering
\includegraphics[width=0.7\linewidth]{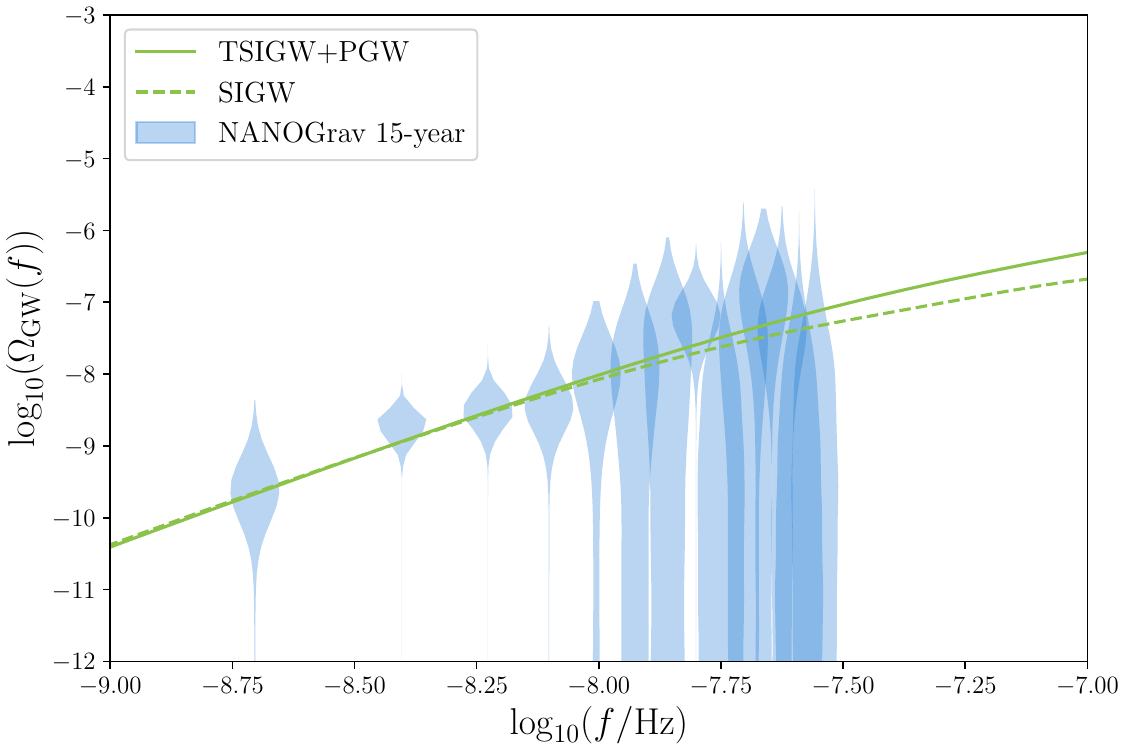}
\caption{The spectra energy density of \acp{PGW}$+$\acp{TSIGW} and \acp{SIGW}. The energy density spectra derived from the free spectrum of the NANOGrav 15-year are shown with blue. The green curves represent the energy density spectra of \acp{GW} with different line styles labeled in the figure. Specifically, the parameters for the green solid line are: $\log_{10}(A_{\zeta})=-1.12$, $\log_{10}(A_h)=-1.39$, $\log_{10}(f_*/\mathrm{Hz})=-6.42$, and $\sigma_*=1.04$; and for the green dashed line, the parameters are: $\log_{10}(A_{\zeta})=-0.44$,  $\log_{10}(f_*/\mathrm{Hz})=-6.85$, and $\sigma_*=0.85$. These parameters are selected based on the median values of the posterior distributions.} \label{fig:violinplot}
\end{figure}

The posterior distributions of different models in Fig.~\ref{fig:bayes} are presented in Fig.~\ref{fig:corner_SIGW} and Fig.~\ref{fig:corner_TSIGW}, where prior distributions for $\log_{10}(f_*/\mathrm{Hz})$, $\log_{10}(A_\zeta)$, $\log_{10}(A_h)$, and $\sigma_*$ are uniform over the intervals $[-10,3]$, $[-3,1]$,$[-3,1]$, and $[0.1,1.5]$, respectively. More precisely, in Fig.~\ref{fig:corner_SIGW}, we have ignored small-scale \acp{PGW}, focusing on scenarios where \acp{SIGW} are the primary contributors to current \ac{PTA} observations and where both second-order \acp{SIGW} and \ac{SMBHB} contribute to \ac{PTA} observations. In Fig.~\ref{fig:corner_TSIGW}, we examine the scenarios where \acp{PGW}$+$\acp{TSIGW} dominate current \ac{PTA} observations and where \acp{PGW}$+$\acp{TSIGW} and \ac{SMBHB} jointly dominate \ac{PTA} observations. Comparing the posterior distributions in Fig.~\ref{fig:corner_SIGW} and Fig.~\ref{fig:corner_TSIGW} reveals that, when considering large-amplitude \acp{PGW} on small scales, the median value of $A_{\zeta}$ derived from current \ac{PTA} observations is significantly reduced. Furthermore, since the total energy density spectrum of \acp{TSIGW} is composed of the four energy density spectra shown in Fig.~\ref{fig:spectrum}, this leads to a degeneracy between the amplitudes of the primordial power spectra $A_{\zeta}$ and $A_h$. It should be noted that the degeneracy between parameters $A_{\zeta}$ and $A_h$ only exists in the log-normal primordial power spectrum with large $\sigma_*$. For a monochromatic primordial power spectrum, the large-amplitude primordial tensor perturbations on small scales only affect the total energy density spectrum of \acp{GW} in the high-frequency region and do not affect the energy density spectrum in the low-frequency region \cite{Chang:2022vlv}. In this scenario, the degeneracy between parameters $A_{\zeta}$ and $A_h$ is not significant. Fig.~\ref{fig:violinplot} illustrates the total energy density spectra and compare the results with those of second-order \acp{SIGW}. Similar to the second-order \acp{SIGW}, the \acp{PGW}$+$\acp{TSIGW} in the case of a log-normal primordial power spectrum also fit well with current \ac{PTA} observational data.

\subsection{Signal-to-noise ratio of LISA}\label{sec:4.2}
\begin{figure*}[htbp]
    \captionsetup{
      justification=raggedright,
      singlelinecheck=true
    }
    \centering
	\subfloat[$\sigma_* = 0.5$, $A_h=0.02$, $f_*=5\times10^{-3}\mathrm{Hz}$ ]{\includegraphics[width=.45\columnwidth]{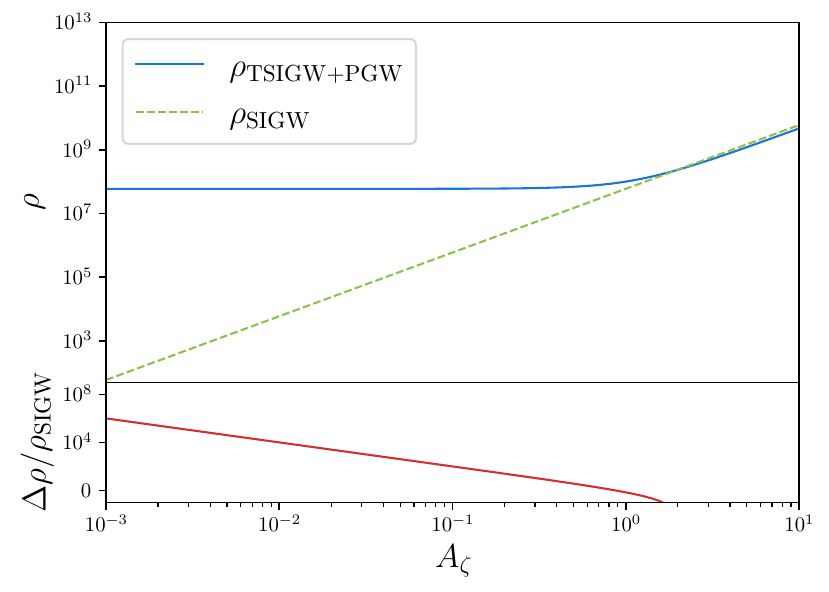}}
        \subfloat[$A_\zeta = 0.1$, $\sigma_* = 0.5$, $f_*=5\times10^{-3}\mathrm{Hz}$ ]{\includegraphics[width=.45\columnwidth]{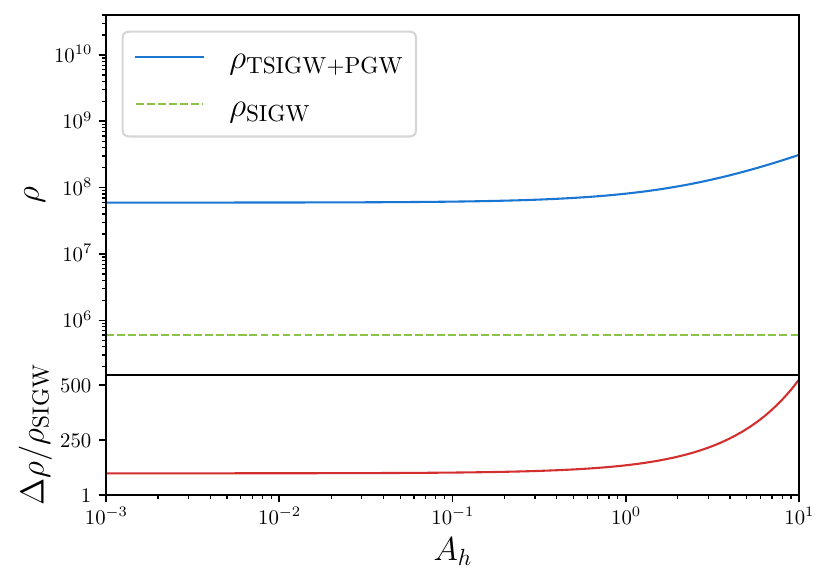}} \\
        \subfloat[$A_\zeta = 0.1$, $\sigma_* = 0.5$, $A_h=0.02$ ]{\includegraphics[width=.45\columnwidth]{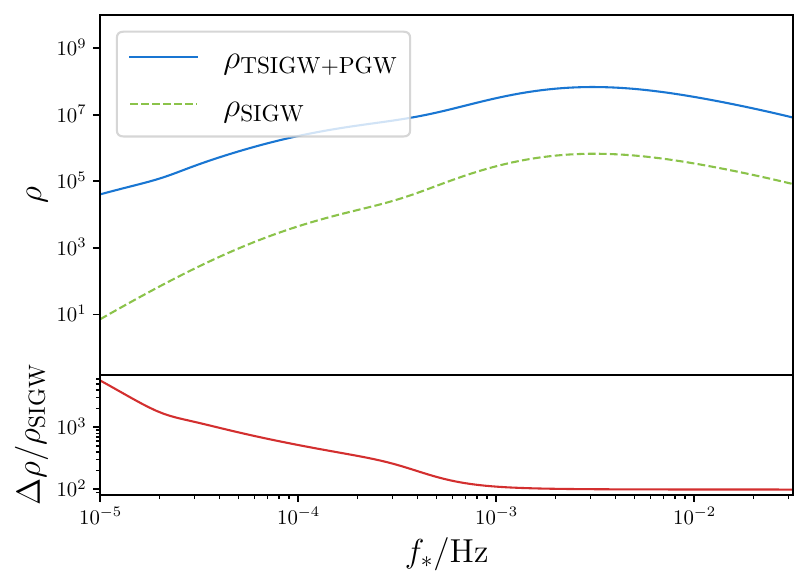}}
        \subfloat[$A_\zeta = 0.1$, $A_h=0.02$, $f_*=5\times10^{-3}\mathrm{Hz}$ ]{\includegraphics[width=.45\columnwidth]{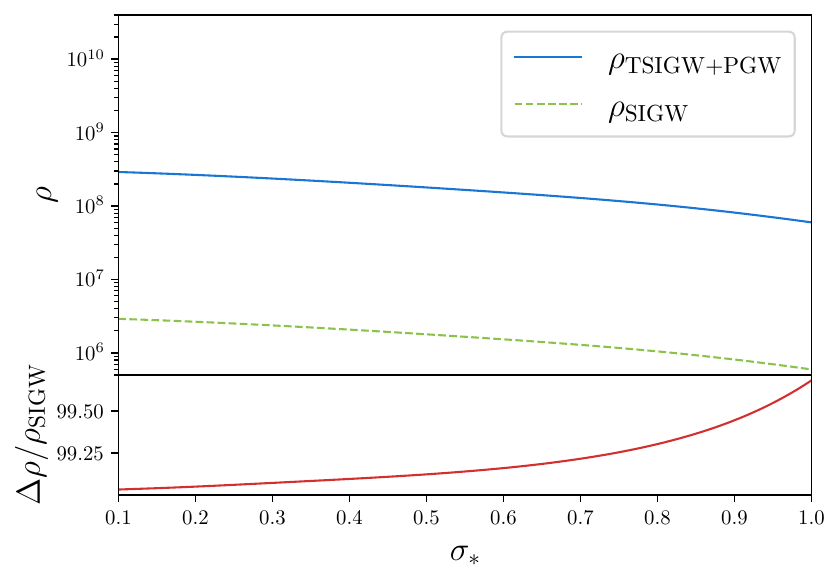}}    
\caption{\label{fig:LISA_fixf} The \ac{SNR} of \ac{LISA} as a function of the parameters in primordial power spectra for both the \acp{SIGW} and \acp{PGW}$+$\acp{TSIGW}. The curves below show the improvement of \ac{SNR} after considering the effects of \acp{PGW}. $\Delta \rho=|\rho_{\mathrm{TSIGW+PGW}}-\rho_{\mathrm{SIGW}}|$. }
\end{figure*}

\begin{figure*}[htbp]
    \captionsetup{
      justification=raggedright,
      singlelinecheck=true
    }
    \centering
	\subfloat[The \ac{SNR} of \ac{LISA} as a function of $A_\zeta$ and $A_h$.\label{fig:heatmap_AAh}]  {\includegraphics[width=.47\columnwidth]{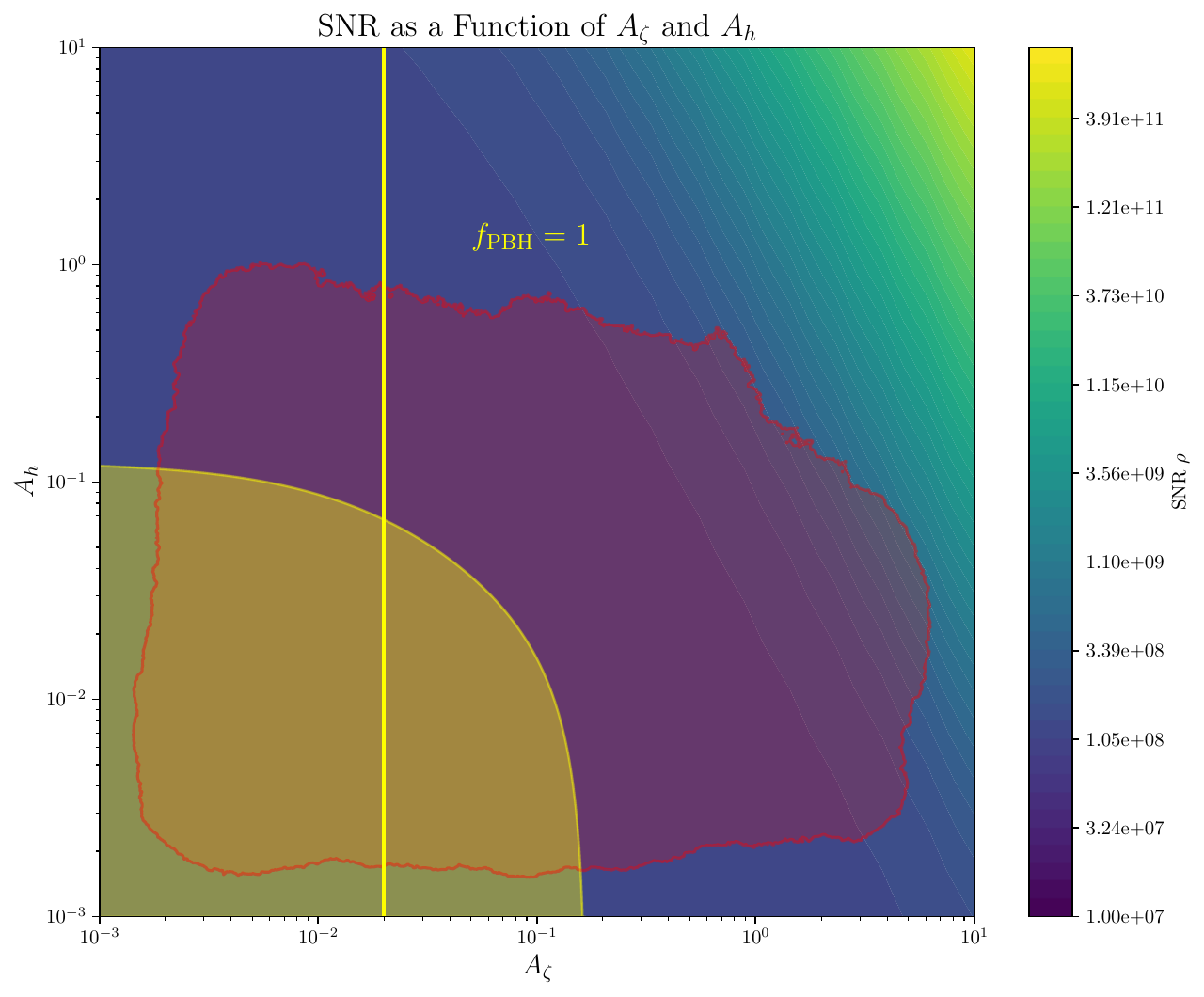}}
    \hspace{0.1cm}
        \subfloat[The \ac{SNR} of \ac{LISA} as a function of $f_*$ and $\sigma_*$.\label{fig:heatmap_fsigma}]{\includegraphics[width=.47\columnwidth]{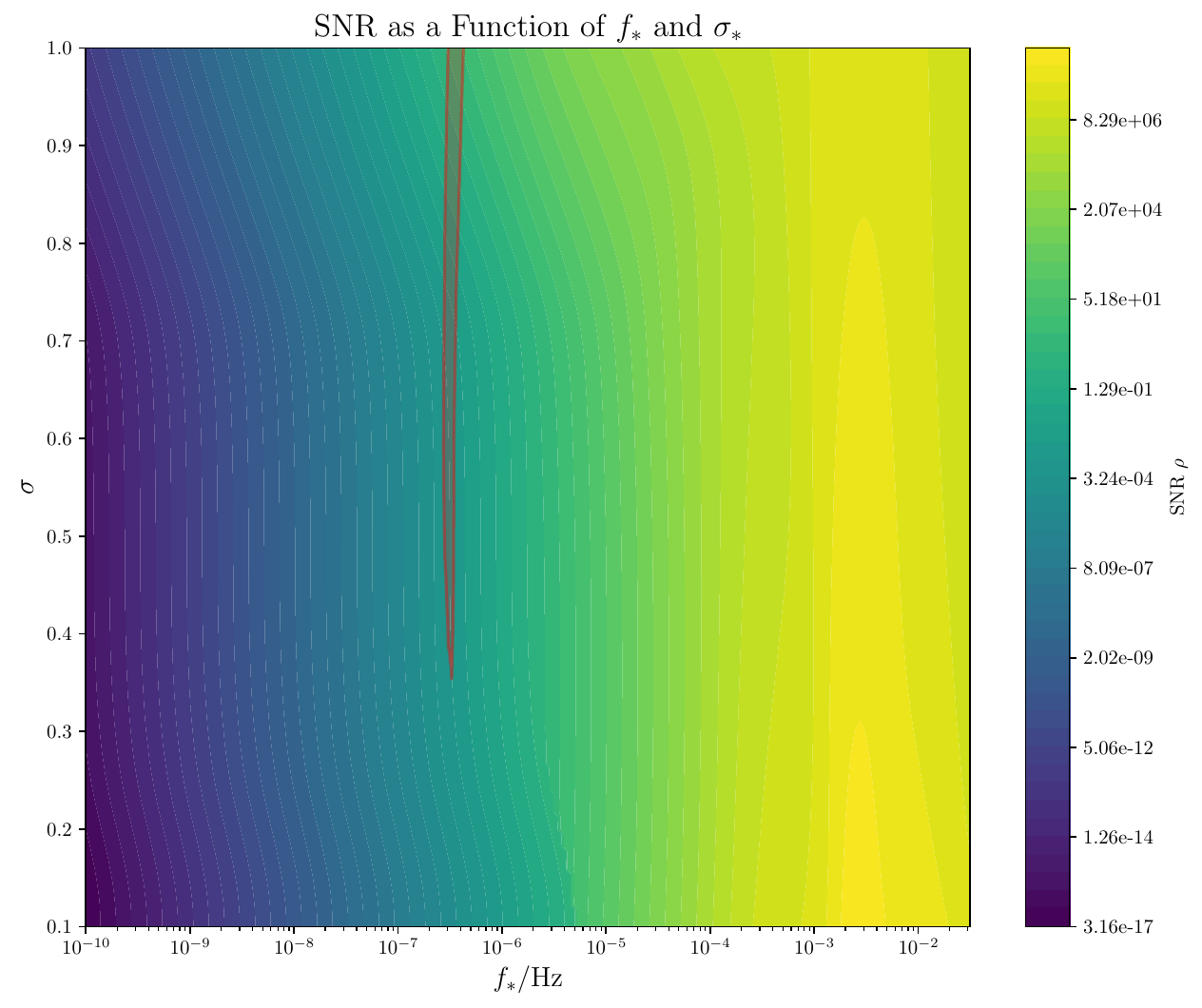}} 
\caption{\textbf{Left panel: }The \ac{SNR} of \ac{LISA} as a function of $A_\zeta$ and $A_h$. The heatmap shows the \ac{SNR} for fixed values of $f_* = 5 \times 10^{-3}$ Hz and $\sigma_* = 0.5$. The red shade represents the $68\%$ credible intervals of the 2D posterior distribution in Fig.~\ref{fig:corner_TSIGW}, obtained through the \ac{KDE} method implemented in the software ChainConsumer \cite{Hinton2016,Hinton2016_software}. The yellow shade shows the constraints from \ac{CMB} and \ac{BAO} when $\sigma_*=0.5$, which are independent of $f_*$. The yellow line represents the curve for $f_*\approx 10^{-8}$ Hz, $\sigma_*=0.5$,  $f_{\mathrm{PBH}}=1$, indicating that all of dark matter is composed of \acp{PBH} \cite{Carr:2020xqk,Carr:2016drx,DeLuca:2019qsy,Iovino:2024tyg,Musco:2020jjb}. \textbf{Right panel: }the \ac{SNR} of \ac{LISA} as a function of $f_*$ and $\sigma_*$. The heatmap shows the \ac{SNR} for fixed values of $A_\zeta = 0.1$ and $A_h = 0.02$. The red shade represents the $68\%$ credible intervals of the 2D posterior distribution in Fig.~\ref{fig:corner_TSIGW}. }
\end{figure*}

In the previous subsections, we investigated the effects of \acp{PGW}$+$\acp{TSIGW} on current \ac{PTA} observations. As discussed in Sec.~\ref{sec:3.0}, large-amplitude \acp{PGW} on small scales significantly affect the total energy density spectrum of induced \acp{GW} in the high-frequency region. Therefore, in this subsection, we discuss the impact of \acp{PGW}+\acp{TSIGW} on the \ac{SNR} of \ac{LISA} to quantitatively assess their effect on the total energy density spectrum in the high-frequency region. The \ac{SNR} of \ac{LISA} can be expressed as \cite{Siemens:2013zla, Robson:2018ifk}
\begin{equation}
    \rho = \sqrt{T}\left[ \int \mathrm{d} f\left(\frac{\bar{\Omega}_{\mathrm{GW},0}(f)}{\Omega_\mathrm{n}(f)}\right)^2\right]^{1/2} \ ,
\end{equation}
where $T$ is the observation time and we set $T=4$ years here. $\Omega_\mathrm{n}(f)=2\pi^2f^3S_n/3H_0^2$, where $H_0$ is the Hubble constant, $S_n$ is the strain noise power spectral density \cite{Robson:2018ifk}. Fig.~\ref{fig:LISA_fixf} shows the \ac{SNR} curves for LISA experiments. $\rho_{\mathrm{SIGW}}$ and $\rho_{\mathrm{TSIGW+PGW}}$ represent the \ac{SNR} for the second-order \acp{SIGW} and \acp{PGW}$+$\acp{TSIGW}, respectively. Fig.~\ref{fig:LISA_fixf} demonstrates that the existence of \acp{PGW} substantially increases \ac{SNR} of \ac{LISA} in the high-frequency range. Moreover, Fig.~\ref{fig:heatmap_AAh} and Fig.~\ref{fig:heatmap_fsigma} illustrate the two-dimensional distribution of the SNR of LISA across various primordial power spectrum parameters. The red areas in Fig.~\ref{fig:heatmap_AAh} and Fig.~\ref{fig:heatmap_fsigma} correspond to the $68\%$ credible intervals of the 2D posterior distribution, determined when TSIGWs dominate PTA observations.
As depicted in Fig.~\ref{fig:LISA_fixf}, with the rise in the amplitudes $A_{\zeta}$ and $A_h$ of the primordial power spectra, the energy density spectrum of \acp{PGW}$+$\acp{TSIGW} will increase, leading to a gradual increase in the \ac{SNR} of \ac{LISA}. In Fig.~\ref{fig:heatmap_fsigma}, the parameter $f_*$ indicates the position of the peak of the log-normal primordial power spectrum in Eq.~(\ref{eq:Pn1}), which also determines the frequency band of the energy density spectrum of \acp{PGW}$+$\acp{TSIGW}. As the parameter $f_*$ increases, the energy density spectrum of \acp{PGW}$+$\acp{TSIGW} will gradually shift entirely into the detection band of LISA, resulting in an increase in \ac{SNR} of \ac{LISA}.

\subsection{Constraints from CMB and BAO}\label{sec:4.3}
Following the discussion in the previous subsection,  \acp{TSIGW}, which are a significant source of the \acp{SGWB}, affect \ac{PTA} observations. Additionally, they can serve as an extra radiation component, thereby influencing the cosmological observations on large scales  \cite{Clarke:2020bil,Zhou:2024yke}. More precisely, the total energy density of \acp{GW} satisfies \cite{Zhou:2024yke,Wang:2023sij,Wright:2024awr,Sui:2024grm}
\begin{eqnarray}\label{eq:rhup}
h^2\rho_{\mathrm{GW}}=\int_{f_{\mathrm{min}}}^{\infty} h^2\Omega_{\mathrm{GW},0}(k) \mathrm{d}\left(\ln k\right) < 2.9\times 10^{-7} 
\end{eqnarray}
at $95\%$ confidence level for \ac{CMB}$+$\ac{BAO} data \cite{Clarke:2020bil}. As shown in the yellow shade in Fig.~\ref{fig:heatmap_AAh}, we present the current \ac{CMB}$+$\ac{BAO} observational data constraints on the parameter space of the amplitude of primordial power spectra. As discussed in Sec.~\ref{sec:4.2}, the red shaded area in Fig.~\ref{fig:heatmap_AAh} represents the parameter space for the amplitude of the primordial power spectrum as determined by PTA observations. The overlap between the red and yellow shaded regions in Fig.~\ref{fig:heatmap_AAh} indicates the constraints on the parameter space for the amplitude of the primordial power spectrum provided by current \ac{CMB}$+$\ac{BAO}$+$\ac{PTA} observations. Furthermore, the constraints on the abundance of \acp{PBH} dictate that the parameter $A_{\zeta}$ should not be greater than the yellow line shown in Fig.~\ref{fig:heatmap_AAh}. In summary, the main results regarding the amplitude of the primordial power spectrum in this paper are as follows:

\tcbset{colback=gray!20, colframe=gray!20, boxrule=0.5mm, arc=0mm, auto outer arc, width=\linewidth} \begin{tcolorbox} 
\textbf{ The overlapping area of the red and yellow shaded regions to the left of the yellow line in Fig.~\ref{fig:heatmap_AAh} indicates the parameter space for the amplitude of the primordial power spectrum permitted by current cosmological observations (\ac{CMB}$+$\ac{BAO}$+$\ac{PTA}$+$\acp{PBH}).}
\end{tcolorbox}

The results show that if \acp{PGW}$+$\acp{TSIGW} dominates the current PTA observations, the amplitude of power spectrum of primordial curvature perturbation is about $10^{-2}$. In this study, we only consider the Gaussian primordial  perturbations. In the case of local-type primordial non-Gaussianity, if we only consider second-order \acp{SIGW} or \acp{TSIGW}, the effects of  local-type  primordial non-Gaussianity do not significantly affect the total energy density spectrum of \acp{GW}. However, when the primordial perturbations on small scales are substantial, the contributions from third-order and higher-order induced \acp{GW} become significant. As indicated in Ref.~\cite{Chang:2023aba}, the cross-correlation function $\langle h^{\lambda,(3)}_{\mathbf{k}}(\eta) h^{\lambda',(2)}_{\mathbf{k}'}(\eta) \rangle$ between second-order \acp{SIGW} and third-order \acp{SIGW} is proportional to the five-point correlation function of the primordial curvature perturbation: $\langle \zeta_{\mathbf{k}-\mathbf{p}}\zeta_{\mathbf{p}-\mathbf{q}}\zeta_{\mathbf{q}} \zeta_{\mathbf{k}'-\mathbf{p}'} \zeta_{\mathbf{p}'}\rangle$. Consequently, $\langle h^{\lambda,(3)}_{\mathbf{k}}(\eta) h^{\lambda',(2)}_{\mathbf{k}'}(\eta) \rangle$ contributes to the energy density spectrum of gravitational waves only in the presence of primordial non-Gaussianity, with its contribution being proportional to $f_{\mathrm{NL}}A_{\zeta}^3+(f_{\mathrm{NL}})^3A_{\zeta}^4+(f_{\mathrm{NL}})^5A_{\zeta}^5$. When the non-Gaussian parameter $f_{\mathrm{NL}}$ is negative, the cross-correlation function $\langle h^{\lambda,(3)}_{\mathbf{k}}(\eta) h^{\lambda',(2)}_{\mathbf{k}'}(\eta) \rangle$ will notably suppress the total energy density spectrum of induced \acp{GW}. Therefore, the conclusion here may be affected by various physical processes during the evolution of the universe, such as the impact of primordial non-Gaussianity \cite{Franciolini:2023pbf,Iovino:2024sgs,Picard:2024ekd}, varying sound speed \cite{Jin:2023wri,Balaji:2023ehk}, interactions between induced gravitational waves and matter \cite{Zhang:2022dgx,Mangilli:2008bw,Yu:2024xmz,Loverde:2022wih,Saga:2014jca,Sui:2024nip}, and higher-order effects \cite{Chang:2023vjk,Chang:2023aba}. Moreover, as depicted in Fig.~\ref{fig:heatmap_fsigma} by the red shaded regions, the \ac{SNR} of \ac{LISA} is quite minimal within the parameter space identified by \ac{PTA} observations. Therefore, the conclusions on \acp{TSIGW} observations in this paper are as follows: 

\tcbset{colback=gray!20, colframe=gray!20, boxrule=0.5mm, arc=0mm, auto outer arc, width=\linewidth} \begin{tcolorbox} 
\textbf{ If the \acp{TSIGW}$+$\acp{PGW}  dominates the \acp{SGWB} observed in the current \ac{PTA} frequency band, its impact on the \acp{SGWB} observations in the future \ac{LISA} frequency band is negligible.}
\end{tcolorbox}

The conclusions above provide a method for identifying different sources of the \acp{SGWB}, namely by observing \acp{SGWB} on different scales and combining other types of observational constraints (such as \acp{PBH}, \ac{CMB}, \ac{BAO} etc.) to jointly constrain or determine the source of the \acp{SGWB}. This method is applicable not only to \acp{TSIGW} but also to other types of \acp{SGWB} generated in the early universe \cite{Ellis:2023oxs}. It should be noted that these results depend on the specific form of the primordial power spectrum, which may lead to different conclusions for different forms. For other forms of the primordial power spectra, we need to use the model-independent results given in Eq.~(\ref{eq:P11})--Eq.~(\ref{eq:Pij}) to recalculate the energy density spectrum of \acp{TSIGW}.

\section{Conclusion and discussion}\label{sec:5.0}
Primordial curvature and tensor perturbations may exhibit significant amplitudes at small scales, potentially inducing higher-order \acp{GW} with observable effects. In this paper, we investigated second-order \acp{TSIGW} during the \ac{RD} era. The source terms of second-order \acp{TSIGW} are divided into three types: scalar-scalar, tensor-scalar, and tensor-tensor sources. When ignoring large-amplitude primordial tensor perturbations at small scales, our results for second-order \acp{TSIGW} align with the results of second-order \acp{SIGW}. We provided explicit expressions for the energy density spectrum of second-order \acp{TSIGW}. Considering a log-normal primordial power spectrum, we utilized current \ac{CMB}$+$\ac{BAO}$+$\ac{PTA}$+$\acp{PBH} observational data to constrain the parameter space of the small-scale primordial power spectrum. Furthermore, we analyzed the impact of different primordial power spectrum parameters on the \ac{SNR} of \ac{LISA} in the high-frequency region. The specific constraints on the parameter space of primordial curvature and tensor perturbations from current observations are shown in the overlapping area of the red and yellow shaded regions to the left of the yellow line in Fig.~\ref{fig:heatmap_AAh}. The presence of large-amplitude primordial tensor perturbations on small scales can prevent the overproduction of \acp{PBH} \cite{Gorji:2023sil}.

The results we presented in Eq.~(\ref{eq:P11})--Eq.~(\ref{eq:Pij}) are applicable to any form of primordial curvature perturbation and primordial tensor perturbation power spectra. Given specific primordial power spectra, we can calculate the corresponding energy density spectrum of second-order \acp{TSIGW}. Notably, when the peaks of the primordial curvature and tensor perturbation power spectra occur at different scales, the second-order \acp{GW} induced by tensor-scalar sources can be very large. This potential "divergence" issue in the energy density spectrum of second-order \acp{TSIGW} has been  studied in Refs.~\cite{Bari:2023rcw,Picard:2023sbz}. Ref.~\cite{Bari:2023rcw} provides a phenomenological method to mitigate such divergence by inserting a correction factor $f(u)=\frac{u^4}{d^4+u^4}$ into Eq.~(\ref{eq:P22}) to "dilute" the potential divergence effect. Understanding the potential "divergence" issue within tensor-scalar sources remains an unresolved question and may be further refined in future research.

In current studies of second-order \acp{TSIGW}, the chosen primordial power spectra are typically parameterized log-normal primordial power spectra. For specific primordial power spectra provided by particular inflation models, we can use Eq.~(\ref{eq:P11})--Eq.~(\ref{eq:Pij}) to calculate the energy density spectrum of second-order \acp{TSIGW} and use the cosmological observations at different scales given in Sec.~\ref{sec:4.0} to constrain the parameter space of these inflation models. Furthermore, as illustrated in Fig.~1, we focus primarily on the one-loop contributions from the two-point correlation function $\langle h^{\lambda,(2)}_{\mathbf{k}} h^{\lambda^{\prime},(2)}_{\mathbf{k}'}\rangle$. The one-loop contributions from the two-point correlation function $\langle h^{\lambda,(3)}_{\mathbf{k}} h^{\lambda^{\prime},(1)}_{\mathbf{k}'}\rangle$, as considered in Ref.~\cite{Chen:2022dah}, and the two-loop contributions similar to the \acp{SIGW} studies \cite{Chang:2023vjk}, have not been systematically investigated. The comprehensive one-loop calculations for \acp{TSIGW}, along with the corresponding two-loop calculations, may be further developed in future research.

\acknowledgments
We thank Dr.~Y.H. Yu for the useful discussions. The work is supported in part by the National Natural Science Foundation of China (NSFC) grants  No.12475075, No. 11935009, No. 12375052,  and No. 12447127.

\bibliography{biblio}

\end{document}